\documentclass[10pt,preprint2]{aastex}             
\usepackage{psfig}

\slugcomment{Revised manuscript submitted \today}

\shortauthors{Strickland \etal}
\shorttitle{Diffuse X-ray emission in the halo of NGC 253}

\lefthead{Strickland \etal}
\righthead{Diffuse X-ray emission in the halo of NGC 253}

\newcommand{\eg}{{\rm e.g.\ }}

\newcommand{\ie}{{\it i.e.\ }}

\newcommand{\etal}{{\rm et al.\thinspace}}

\newcommand{\cm}{{\rm\thinspace cm}}

\newcommand{\pcc}{\hbox{$\cm^{-3}\,$}}
\newcommand{\s}{{\rm\thinspace s}}
\newcommand{\yr}{{\rm\thinspace yr}}
\newcommand{\erg}{{\rm\thinspace erg}}
\newcommand{\ps}{\hbox{\s$^{-1}\,$}}
\newcommand{\pyr}{\hbox{\yr$^{-1}$}}
\newcommand{\ergps}{\hbox{$\erg\s^{-1}\,$}}

\newcommand{\pcmsq}{\hbox{$\cm^{-2}\,$} }

\newcommand{\halpha}{H$\alpha$}

\newcommand{\K}{{\rm K}}
\newcommand{\hi}{H{\sc i}}

\newcommand{\nii}{[N{\sc ii}]}

\newcommand{\nH}{\hbox{$N_{\rm H}$}}

\newcommand{\pc}{{\rm\thinspace pc}}
\newcommand{\kpc}{{\rm\thinspace kpc}}

\newcommand{\keV}{{\rm\thinspace keV}}

\newcommand{\Lsol}{\hbox{$\thinspace L_{\sun}$}}
\newcommand{\Msol}{\hbox{$\thinspace M_{\sun}$}}

\begin{document}

\title{Chandra observations of NGC 253. II: On the origin of
	diffuse X-ray emission in the halos of starburst galaxies}


\author{David K. Strickland,\altaffilmark{1,2} 
	Timothy M. Heckman,\altaffilmark{2}
	Kimberly A. Weaver,\altaffilmark{3}
	Charles G. Hoopes,\altaffilmark{2} and
	Michael Dahlem\altaffilmark{4}}

\altaffiltext{1}{{\it Chandra} Fellow.}

\altaffiltext{2}{Department of Physics and Astronomy, 
	The Johns Hopkins University,
	3400 North Charles Street, Baltimore, MD 21218}

\altaffiltext{3}{NASA/Goddard Space Flight Center, 
	Code 662, Greenbelt, Maryland 20771}

\altaffiltext{4}{European Southern Observatory, 
	Casilla 19001, Santiago 19, Chile}

\begin{abstract}
We present a detailed case study of the diffuse X-ray and \halpha~emission
in the halo of NGC 253, a nearby edge-on starburst galaxy driving a
galactic superwind. The arcsecond spatial resolution
of the ACIS imaging spectroscope on the {\it Chandra} X-ray Observatory
allows us to study the spatial and spectral properties of the diffuse
X-ray emitting plasma, at a height of between 3 and 9 kpc above the disk
in the northern halo of NGC 253, with greatly superior
spatial and spectral resolution compared to previous X-ray instruments.
We find statistically significant structure within the diffuse emission
on angular scales down to $\sim 10\arcsec$ ($\sim 130$ pc), and place limits
on the luminosity of any X-ray-emitting ``clouds'' on smaller scales.
There is no statistically significant evidence for any spatial
variation in the spectral properties of the diffuse emission
over scales from several $\sim 400$ pc to $\sim 3$ kpc.
The spectrum of the diffuse X-ray emission is clearly thermal, although
with the higher spectral resolution and sensitivity of {\it Chandra}
it is clear that current simple spectral models do not provide
a physically meaningful description of the spectrum. In particular, the
fitted metal abundances are unphysically low.
There is no convincing evidence for diffuse X-ray emission at
energies above 2 keV in the halo.

We show that the X-shaped soft X-ray morphology of
the superwind previously revealed by {\it ROSAT} is matched
by very similar X-shaped \halpha~emission, extending at least 8 kpc
above the plane of the galaxy. In the northern halo the X-ray
emission appears to lie slightly interior to
 the boundary marked by the \halpha~emission.
The total 0.3 -- 2.0 keV energy band
 X-ray luminosity of the northern halo $L_{\rm X} \sim 5 
\times 10^{38} \ergps$, is very similar
to the halo \halpha~luminosity of $L_{H\alpha} \sim 4 \times 10^{38} \ergps$, 
both of which are a small fraction of the estimated wind energy injection rate
of $\sim 10^{42} \ergps$ from supernovae in the starburst.
We show that there are a variety of models that can simultaneously 
explain spatially-correlated X-ray and \halpha~emission in the
halos of starburst galaxies, although the physical origin of the
various emission components can be very different in different
models.
These findings indicate that the physical origin of 
the X-ray-emitting million-degree plasma in superwinds is closely
linked to the presence of much cooler and denser $T \sim 10^{4}$ gas,
not only within the central kpc regions of starbursts, but also
on $\sim 10$ kpc-scales within the halos of these galaxies.
\end{abstract}

\keywords{ISM: jets and outflows --- ISM: bubbles ---
galaxies: halos --- galaxies: individual (NGC 253) 
--- galaxies: starburst --- X-rays: galaxies}

\section{Introduction}
\label{sec:introduction}

NGC 253 is an archetypal starburst galaxy at a distance
of only $D=2.6$ Mpc \citep{puche88}, and is the closest
edge-on IR luminous starburst galaxy
($L_{\rm FIR} = 1.7 \times 10^{10} \Lsol$, \citet{radovich_isophot}).
As such it represents the
best local laboratory to study the effects of stellar
feedback from strong star formation on the gaseous interstellar
medium (ISM), processes which are intimately connected to
galaxy formation and galaxy evolution.

Potentially the most significant form of stellar feedback is
the driving of high-velocity, galactic-scale, multi-phase outflows ---
``superwinds'' (\citet{ham90}; \citet{hla93}; \citet{bland_hawthorn95}).
These supernova-driven
outflows are ubiquitous in all local starburst galaxies 
\citep{lehnert_heckman96},
and there is mounting evidence that they are extremely
common in high red-shift galaxies \citep{pettini2001}.

Among the many implications of superwinds, their
possible role in enriching and heating the intergalactic medium (IGM)
is currently attracting much attention (\eg \citet{nath97};
\citet{ferrara2000}; \citet{lloyd_davies}. For an up-to-date
review see \citet{heckman_2001_review}).

X-ray observations of superwinds can in principle tell us much
about the energetics and composition of these outflows, and hence
the transport of energy and metal-enriched gas into the IGM. Superwinds
are driven by the thermal pressure of hot (up to $T \sim 10^{8}\K$)
thermalized SN-ejecta,
gas which is expected to be highly metal-enriched. Adiabatic
expansion cools the gas, converting the thermal energy into bulk motion
at velocities up to $v \sim 3000$ km/s. Gas over-run by the expanding
wind, either within the disk or the halo, may be shock-heated to 
million degree temperatures and itself become a source of thermal
X-ray emission \citep{cc}. Other processes, such as thermal conduction or
turbulent mixing between cool dense gas and surrounding hot regions,
can also lead to X-ray-emitting interface regions \citep{weaver77}.

The hydrodynamical complexity of these outflows (see \eg \citep{suchkov94};
\citet{ss2000}), and in particular the
range of possible physical mechanisms generating X-ray emission in starburst
galaxies, has complicated the interpretation of existing
X-ray observations of starburst galaxies with superwinds 
with the low spatial-resolution instruments that were available
before the launch of the {\it Chandra} X-ray Observatory in 1999
(see the discussions
in \citet{dwh98}; \citet{ss2000}; \citet{whd}; \citet{strickland2000}).

In this paper we present a detailed case study of the diffuse X-ray emission
from the superwind that extends $\sim 10 \kpc$ into the northern halo of 
NGC 253 with {\it Chandra}. Our aim is
to use this data, along with deep ground-based optical \halpha~imaging,
to investigate the physical origin of the extended soft thermal X-ray emission
seen in a {\em typical} starburst-driven superwind. 

The fundamental advantage {\it Chandra} offers over any other
previous or currently-existing instrument is $\sim1\arcsec$-spatial
resolution ($1\arcsec \approx 13$ pc at the distance of NGC 253), 
$\sim 10$ times better spatial
resolution than {\it XMM-Newton}. This allows us to robustly 
remove emission from unrelated point sources and resolve out
any spatial variations in the spectral properties of the diffuse emission.
With {\it Chandra} we can obtain {\em spatially-resolved
spectroscopy of diffuse emission alone}. In contrast, {\it ASCA}, 
{\it BeppoSAX} or combined {\it ROSAT} 
PSPC and {\it ASCA} spectroscopy invariably 
covered the entire galaxy, including all point source and possible
AGN emission (\eg \citet{persic98}; DWH98). 
In addition, the spectral resolution of {\it Chandra}
is comparable to spectral studies with {\it ASCA} \citep{ptak97}
and significantly better than that of the {\it ROSAT} PSPC-based
studies of superwinds (\citet{read94}; \citet{rps97}; \citet{pietsch2000}).

This work follows on from our
{\it Chandra} study of the diffuse X-ray emission in the central kiloparsec
of NGC 253, which explored the brighter inner-most part of the superwind
\citep{strickland2000}.

\begin{figure*}[!ht]
\epsscale{2.0}
\plotone{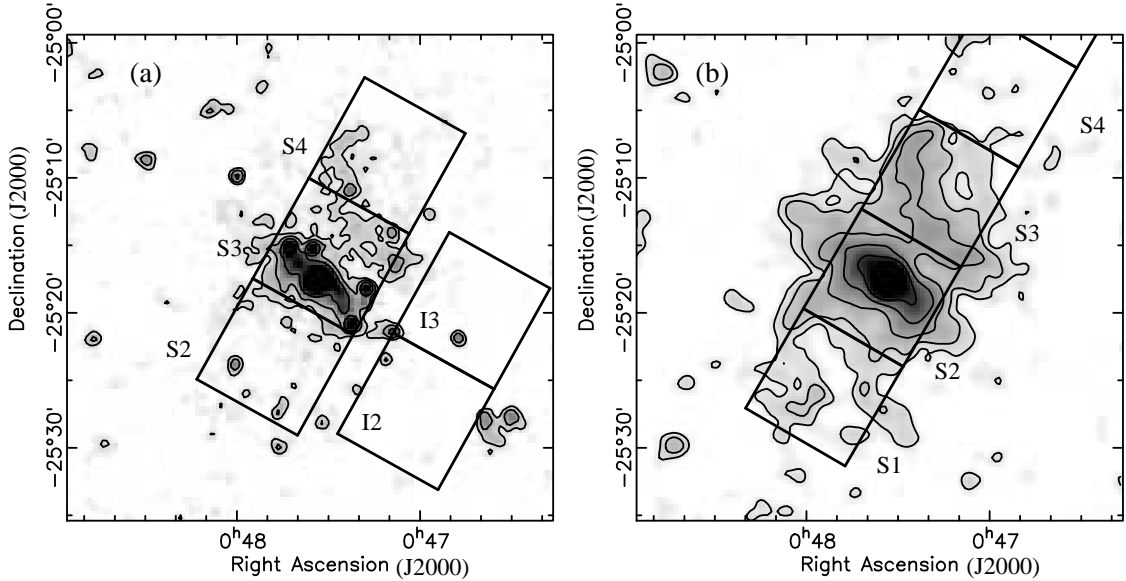}
  \caption{{\it ROSAT} PSPC images of NGC 253, showing the position of 
    the ACIS chips in the two {\it Chandra} observations with respect to the
    large scale diffuse X-ray emission from the superwind.
    (a) ACIS chip orientation in the $\sim 13$ks observation 
    ({\it Chandra} ObsID 969), overlaid on the 0.1 -- 2.0 keV 
    {\it ROSAT} PSPC image. The PSPC image has been smoothed with a
    FWHM = $35\arcsec$ Gaussian mask, and is shown on a logarithmic
    intensity scale between $10^{-4}$ and $10^{-2}$ counts s$^{-1}$
    arcmin$^{-2}$. Contours are shown at a surface brightness levels of
    $1.6, 3.2$ and 6.4 $\times10^{-3}$ counts s$^{-1}$ arcmin$^{-2}$.
    (b) ACIS chip orientation in the $\sim 43$ks observation
    ({\it Chandra} ObsID 790), overlaid on the 0.1 -- 2.0 keV 
    {\it ROSAT} PSPC image from which all the {\it ROSAT}-detected 
    point sources  have been removed (except in the nucleus of the galaxy)
    and the resulting gaps in the data interpolated over. This image
    provides a better representation of the large-scale diffuse emission.
    The image has been smoothed with a FWHM = $70\arcsec$ Gaussian mask,
    and is shown using the same logarithmic intensity scale as in
    panel (a). Contours are shown at surface brightness levels of
    $0.4, 0.8, 1.6$ and 3.2 $\times10^{-3}$ counts s$^{-1}$ arcmin$^{-2}$.
  }
  \label{fig:pspc_chips}
\end{figure*}


\section{Observations and data analysis}
\label{sec:data_analysis}

\subsection{X-ray observations}
\label{sec:data:observations}

Guest observer observations of NGC 253, using the 
AXAF CCD Imaging Spectrometer (ACIS\footnote{For more information on
the capabilities of {\it Chandra} and ACIS 
see the {\it Chandra} Observatory Guide 
\url{http://asc.harvard.edu/udocs/docs/docs.html}, 
section ``Observatory Guide'',``ACIS''}) 
on board the {\it Chandra}
X-ray Observatory, were obtained on 1999 December 16 ({\it Chandra} 
ObsID 969) and December 27 (ObsID 790), resulting
in a total exposure time on 
the S3 chip (prior to the removal of background flares
described in \S~\ref{sec:data:filtering}) of
13987 and 43522 seconds respectively. 
The back-illuminated S3 CCD was at the focal point of the telescope,
as it is more sensitive to the low energy X-ray photons from the
superwind than the front illuminated ACIS chips.
The observations were positioned so as to place the disk of the galaxy,
and the starburst nucleus, within the S3 chip in the shorter exposure,
and the northern lobe of the superwind in the S3 chip in the longer 
exposure.
Figure~\ref{fig:pspc_chips} shows the orientation and position of
the ACIS chips on the sky, overlaid on a 0.1 -- 2.0 keV energy band image
of the {\it ROSAT} PSPC\footnote{The {\it ROSAT} PSPC data was obtained
from the HEASARC archive, and reduced using the Extended Source
Analysis Software created by Steve Snowden.} data. 

\begin{figure*}[!ht]
\epsscale{2.0}
\plotone{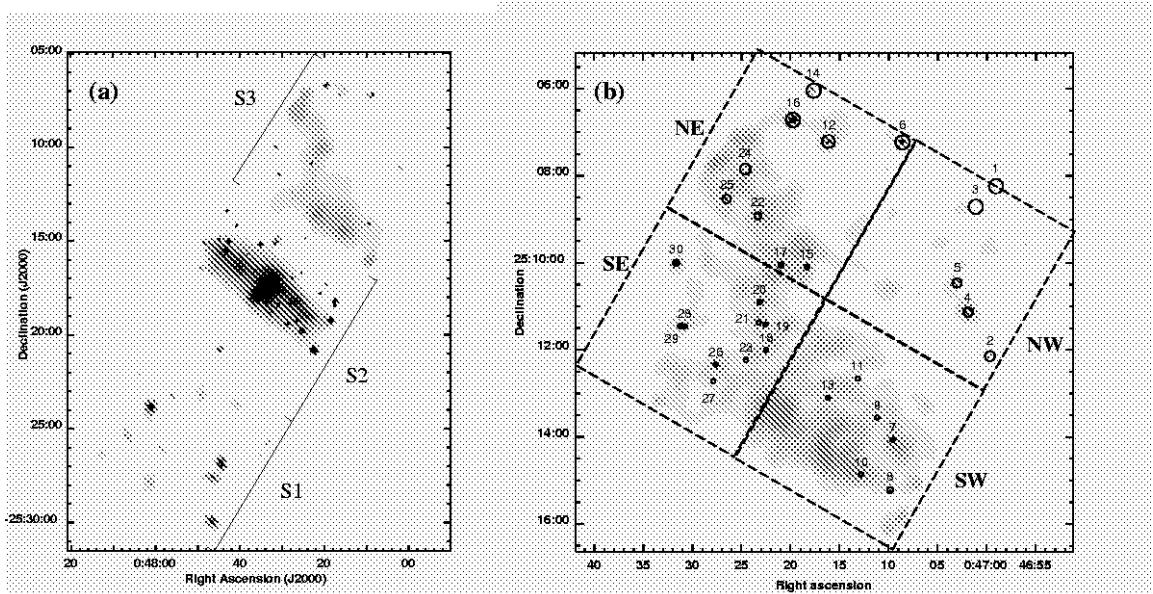}
  \caption{(a) An adaptively smoothed 0.3 -- 8.0 keV image of ACIS
  chips S1, S2 \& S3 from observation ObsID 790.
  (b) Adaptively-smoothed 0.3 -- 8.0 keV image of the northern halo region
  on the S3 chip, showing the regions chosen for diffuse emission
  analysis (Table~\ref{tab:halo_region}) along with all point sources
  with $S/N \ge 2$ (Table~\ref{tab:point_sources}).
  Both images are shown on a logarithmic intensity scale, and have been
  smoothed so that features are locally significant at $\ge 5\sigma$ in
  panel (a) and significant at $\ge 3\sigma$ in panel (b).
  }
  \label{fig:allchips}
\end{figure*}

Compared to the {\it ROSAT} PSPC (the only previous instrument able to attempt
spatially-resolved spectroscopy on diffuse 
emission in the halos of superwinds)
observations of NGC 253, our ACIS observations
have count rates $\sim 3$ times higher and twice the total exposure 
length,
in addition to covering a much wider energy range at $\sim 30$ times
the spatial resolution. 

The ACIS instrument is an imaging CCD spectrometer, comprising
two back-illuminated CCD chips (chips S1 \& S3, nominal in-flight
spectral resolution of $\sim 0.12\keV$ at an photon energy of 1.5 keV,
at the CCD temperature of $-110\degr$ C at which the observations
of NGC 253 were taken),
and eight front-illuminated CCD chips (these chips were damaged by
cosmic rays. Their degraded in-flight
spectral resolution ranges from $0.09\keV$ to
$0.18\keV$ at an photon energy of 1.5 keV,
and depends on the location within the chip).
Each chip is 1024 pixels square, with a plate scale of $0\farcs492$
per pixel. The spatial resolution of the instrument is $\approx 0\farcs8$
(FWHM) on the optical axis with the ACIS-S array in the focal plane,
and degrades to $\sim 6\arcsec$ at an off axis angle of $6 \arcmin$.

\begin{deluxetable}{lccccccccc}
 \tabletypesize{\scriptsize}%
\tablecolumns{9} 
\tablewidth{0pc} 
\tablecaption{Spatial regions used within NGC 253 and its halo
	\label{tab:halo_region}} 
\tablehead{ 
\colhead{Region} & \colhead{$\alpha$ (J2000)} 
	& \colhead{$\delta$ (J2000)}
	& \colhead{Roll}
	& \colhead{Angular size}
	& \colhead{Physical size}
	& \multicolumn{2}{c}{Area} 
	& \multicolumn{2}{c}{Distance from nucleus} \\
\colhead{} & \colhead{(h m s)} 
	& \colhead{($\degr$ $\arcmin$ $\arcsec$)}
	& \colhead{($\deg$)}
	& \colhead{($\arcsec \times \arcsec$)} & \colhead{(kpc $\times$ kpc)}
	& \multicolumn{2}{c}{(arcmin$^{2}$)}
	& \colhead{($\arcsec$)} & \colhead{(kpc)} \\
\colhead{(1)} & \colhead{(2)}
	& \colhead{(3)}
	& \colhead{(4)}
	& \colhead{(5)} & \colhead{(6)}
	& \colhead{(7)} & \colhead{(8)} 
	& \colhead{(9)} & \colhead{(10)}
	}
\startdata
Halo & 00 47 16.43 & -25 10 49.49 
	& 330 & $503.8^{2}$ 
	& $6.35 \times 6.35$ & 70.506 & 69.713 & 449.2 & 5.66 \\
NE halo & 00 47 19.92 & -25 07 57.90 
	& 330 & $251.9^{2}$ 
	& $3.18 \times 3.18$ & 17.627 & 17.187 & 587.4 & 7.40 \\
NW halo & 00 47 03.74 & -25 10 03.39 
	& 330 & $251.9^{2}$ 
	& $3.18 \times 3.18$ & 17.627 & 17.391 & 589.6 & 7.43 \\
SE halo & 00 47 29.17 & -25 11 35.29 
	& 330 & $251.9^{2}$ 
	& $3.18 \times 3.18$ & 17.627 & 17.545 & 346.1 & 4.36 \\
SW halo & 00 47 12.99 & -25 13 43.00
	& 330 & $251.9^{2}$ 
	& $3.18 \times 3.18$ & 17.627 & 17.579 & 347.6 & 4.38 \\
Disk\tablenotemark{a}& 00 47 33.12 & -25 17 17.57
	& 330 & $503.8 \times 251.9$ 
	& $6.35 \times 3.18$ & 35.252 & 31.827 & 0.0 & 0.0 \\
\enddata 
\tablecomments{Columns (2) \& (3): Coordinates of the center of the region.
	(4): {\it Chandra} observation roll angle. 
	(5 \& 6): Projected size of the region in arcsec (5) and kpc (6)
        within the plane of the sky,
        assuming a distance of 2.6 Mpc to NGC 253.
	(7): Total geometrical area of the region.
	(8): Geometrical area remaining after point source removal.
	(9 \& 10): Projected distance from nucleus (the brightest
	radio source in \citet{ua97}) to the center
	of the region in arcsec (9) and kpc (10).}
\tablenotetext{a}{In addition to excluding all point sources, 
  we used a rectangular region of total 
  length $180\arcsec$ along the minor axis
  of the galaxy by $60\arcsec$ wide, centered on the brightest nuclear 
  radio source, 
  to exclude the bright diffuse emission from nuclear starburst region and the 
  nuclear outflow cones.}
\end{deluxetable}

\begin{figure*}[!ht]
\epsscale{2.0}
\plotone{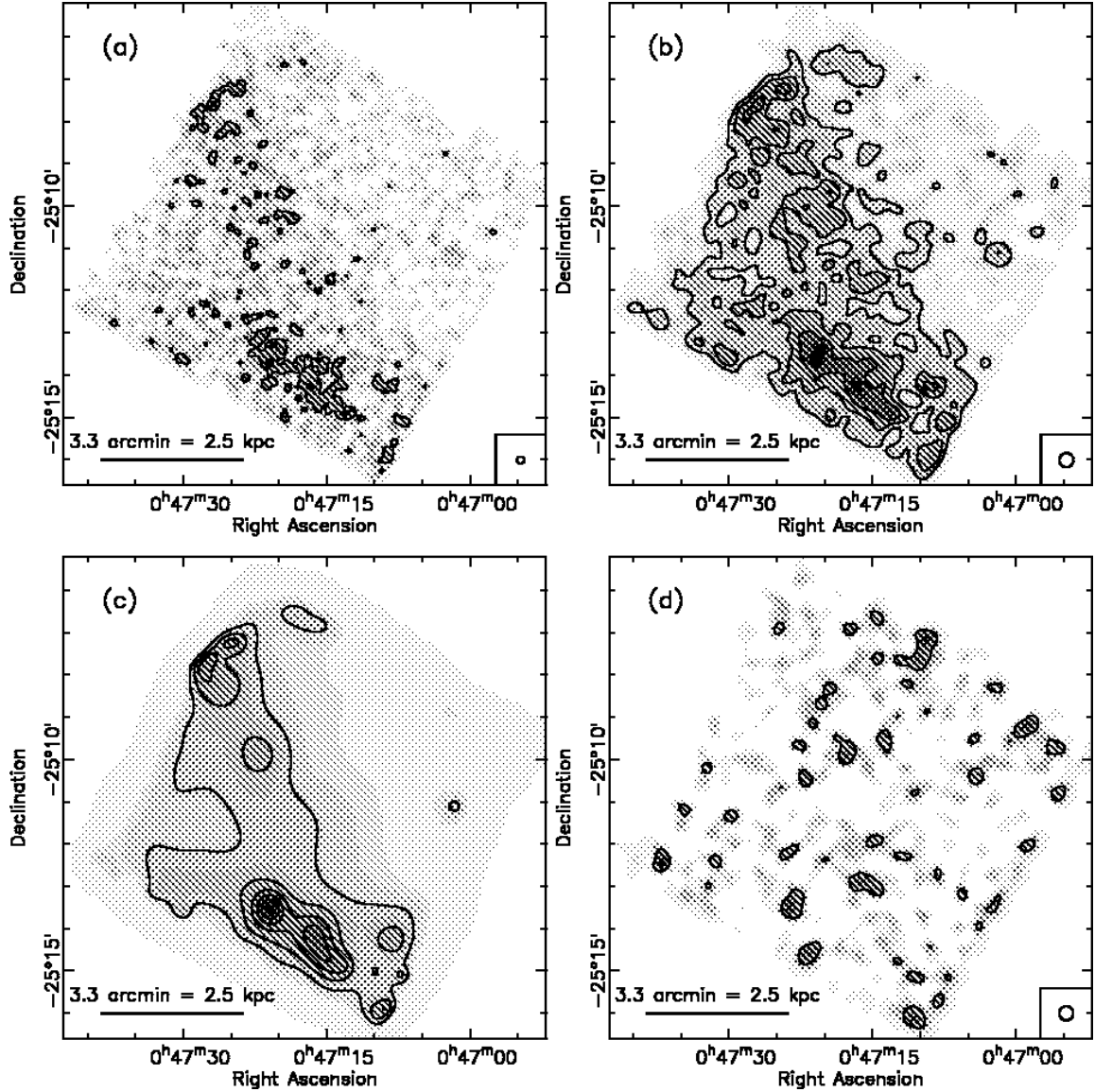}
  \caption{\scriptsize {\it Chandra} ACIS-S3 images of the 
  diffuse X-ray emission
  in the northern halo. Point sources and background emission have been
  subtracted from the data. Panels (a) to (c) show soft X-ray emission
  in the 0.3 -- 2.0 keV energy band, while (d) is a 2.0 -- 8.0 keV
  energy band hard X-ray image. All images use a linear intensity scale
  from zero surface brightness to the maximum surface brightness found in
  that image.
  (a) The soft diffuse emission smoothed with a $FWHM = 10\arcsec$ Gaussian
  mask (the diameter of open circle at the bottom right of the image
  shows the FWHM of the mask). Contours are shown at signal-to-noise values
  of 2 and 3 above the background in a $10\arcsec$-diameter 
  aperture (see Table~\ref{tab:signif_surf} for
  the relationship between surface brightness and signal-to-noise ratio).
  (b) As (a), except smoothed with a $FWHM = 20\arcsec$ Gaussian mask
  to emphasize the larger-scale diffuse emission. Contours are shown
  at signal-to-noise ratios of 2, 3, 4 \& 5 within a $20\arcsec$-diameter
  aperture.
  (c) An adaptively-smoothed image of the diffuse soft X-ray emission.
  Structures visible in this image are significant at the $3\sigma$ level
  above the local diffuse emission and background. Contours are shown
  at  0.9, 1.35, 1.80, 2.25 \& $2.70 \times 10^{-6}$ counts s$^{-1}$ 
  arcsec$^{-2}$.
  (d) The hard X-ray emission remaining after point source and background
  subtraction, smoothed with a  $FWHM = 20\arcsec$ Gaussian. 
  Contours are shown
  at signal-to-noise ratios of 1 \& 2 within a $20\arcsec$-diameter
  aperture. There does not appear to be any significant detection of
  diffuse hard X-ray emission from the halo.
  }
  \label{fig:halo_diffuse}
\end{figure*}

The properties of the diffuse emission
in the northern halo of NGC 253 reported in this paper are based 
exclusively on S3 chip data from the longer observation. 
The halo region does fall within the S4 chip in
the shorter {\it Chandra} observation, 
but inspection of the data
reveals that it adds nothing to what can be obtained from the
targeted S3 observation of the halo.
The combination of a shorter exposure time, 
a larger off-axis angle leading to
a lower spatial resolution, and a 
CCD readout problem with the S4 chip that requires additional
event filtering, all conspire to make the S4 chip data on the halo
superfluous compared to the longer observation.

The southern halo of NGC 253 is partially covered by the
back-illuminated S1 chip in our longer observation. As the center
of the S1 chip is $\sim 14\arcmin$ off-axis, the point spread
function (PSF) of the telescope is large\footnote{The
50\%-energy enclosed radius of the {\it Chandra} PSF varies
between $0\farcs4$ and $3\farcs5$ over the S3 chip, and grows
to between $7\arcsec$ to $21\arcsec$ within the S1 chip.} and comparable
in size to the PSF of {\it XMM-Newton} or the {\it ROSAT} PSPC, making it
difficult to identify and remove point sources from the data.
The diffuse emission in this region is also 
fainter than in the northern halo (as can be seen
in Figs.~\ref{fig:pspc_chips} \& \ref{fig:allchips}), 
and analysis of the {\it Chandra}
data reveals no significant structure to the diffuse emission.
For these reasons we have chosen to concentrate upon
the higher quality data from the northern halo.

The data was reduced and analyzed using {\sc Ciao} (version 2.1),
{\sc Heasoft} (version 5.0.2) and {\sc Xspec} (version 11.0.1aj).
The latest {\it Chandra} calibration files available for this
observation were used ({\sc Caldb} version 2.3). The known
problems with these calibration files are discussed in 
\S~\ref{sec:data:calibration}.
Our reduction and processing followed the guide lines released by the 
CXC and the ACIS instrument 
team\footnote{The {\sc Ciao} (v2.1) Science Threads
can be found at http://asc.harvard.edu/ciao/documents\_threads.html,
the ACIS instrument team's recipes 
at http://www.astro.psu.edu/xray/acis/recipes/.}.

\subsubsection{Background images and spectra}

We have used the background event data sets provided by the CXC
to obtain background images and spectra for the S3 chip. We can not
obtain a local estimate of the background from our data alone as
the diffuse emission in the halo of NGC 253 fills the entire S3 chip.
The background data set used provided a total exposure in the S3 chip
of 114886 s, so the statistical uncertainties in the background
counts are significantly reduced compared to using a local
background estimate. 

The disadvantage
of using data taken at different times and along different sight lines
is that one must worry about systematic differences between the 
background in our observation and that in the datasets used to create
the blank-sky background data. We believe this systematic effect is
only significant at photon energies $E > 2$ keV, for the reasons we
discuss later.

\subsubsection{Time filtering and flare removal}
\label{sec:data:filtering}

We filtered out periods of enhanced total count rate (``flares'')
due to low energy protons interacting with the detector. 
We identified and excluded
periods of higher (or lower) than average count rate by creating
a light curve of all events falling with the S3 chip and using
two iterations of a $5\sigma$-clipping technique to identify outliers
and obtain an accurate measurement of the mean ``quiescent'' count
rate. Following the advice of Maxim Markevitch (Markevitch 2000,
private communication) we binned the light curve into 258.28 second-long
bins, which corresponds to exactly 80 CCD exposures.
This time filtering resulted in a total useful exposure of 
39588.3 s, a loss of 9.0\% of the data. Inspection of the resulting
filtered light curve showed all obvious broad-energy-band flaring
had been removed.

\subsubsection{Point source identification}

The standard source searching methods provided as part of {\sc Ciao}
gave unsatisfactory results when used on the mix of point sources
and diffuse emission we find in these observations. Large numbers of
possibly spurious sources were identified at low, but still sensible,
signal-to-noise ratios by both the sliding box-based {\sc Celldetect}
and the wavelet-based {\sc Wavdetect} source searching programs,
(different sources being found depending on which method was used).
These extra sources are found preferentially in regions where 
the diffuse X-ray emission is brightest, suggesting that 
these excess sources were mis-identification of diffuse emission
as point sources. To obtain robust point source identifications
we used the scheme described in Appendix~\ref{sec:appendix:source_searching}.

\subsubsection{Spectral fitting}
\label{sec:data:spectral}

ACIS spectra were extracted using Pulse Invariant (PI) 
data values, in order to
account for the spatial variations in gain between different
chip nodes. In order to use $\chi^{2}$ as the fit statistic 
the spectra were binned to give a minimum of 10
counts per bin after background subtraction.

Spectral responses were created using the latest calibration files
available at the time of writing
 for observations taken with the CCD temperature of
-110 $\degr C$ ({\sc Caldb} version 2.3), 
following the example procedures provided
by the CXC.

\subsubsection{Calibration uncertainties}
\label{sec:data:calibration}

The calibration of the ACIS instrument used at the time of writing
was the best available, but is known to be deficient in a number
of ways. 
\begin{enumerate}
\item {\bf Low energy effective area:}
  The low energy effective area of the S3 chip is uncertain.
  This will impact fitting absorbing hydrogen columns, which in turn
  affects the inferred absorption-corrected fluxes and emission integrals.
\item {\bf Energy scale below 1 keV:} 
  The energy scale of the ACIS S3 chip is believed to be
  $\sim 20$ eV too low at energies below 1 keV.
  We have taken this into account by fitting for the 
  energy offset using the {\sc Gain} command in {\sc Xspec}, although we
  stress this is method of correction can only be considered
  an approximation.
\item {\bf Energy resolution:} The width and shape of line features
   in the current spectral response differ from those
   observed. In particular the spectral responses appear to systematically
   underestimate the width of spectral features by up to 40\%.
\end{enumerate}

These will affect spectral fitting
of the data, although not to a degree that precludes any 
analysis of the data. We will discuss the effect these calibration problems
have on our spectral analysis in \S~\ref{sec:results:spectral}.
We wish to emphasize that these calibration uncertainties have
negligible effect on the image-based analysis of the data.

\subsection{Optical $H\alpha$ and R-band imaging}

Two sets of ground-based optical imaging of NGC 253 were used to
investigate whether NGC 253 has a large scale \halpha~emitting halo,
and to investigate the absolute astrometric accuracy of the
{\it Chandra} data.

The first data set is a mosaic
of previously unpublished R-band and \halpha~images taken by us
with the 4m telescope at CTIO, which only partially 
cover the region of these {\it Chandra} 
observations.

The second set of images were 
taken with the 0.6m Curtis/Schmidt telescope at CTIO, taken as part of a
survey of the diffuse ionized gas in Sculptor group galaxies 
\citep{hoopes96}. 
These images are less sensitive than the 4m imaging, but cover a 
larger field of view and form the basis of our X-ray/\halpha~comparison
study described in \S~\ref{sec:results:halpha}.

The narrow-band \halpha~filter used  has a peak transmission
at 6570~\AA~and FWHM of 68~\AA. The telescope optics cause the filter
passband to be blue-shifted by $\sim 10$~\AA. Combined with a 5~\AA
 ~red-shift due to the systematic velocity of NGC 253, this places 
the \halpha~line and the weak \nii~6548~\AA~line near the peak of the
filter transmission, and the stronger \nii~6584 \AA~line at $\sim
50$\% of the peak filter transmission. Continuum subtraction was performed
using images from a 77~\AA~wide continuum filter centered at 6649~\AA.

We flux-calibrated the resulting continuum-subtracted line image
using the  spectrophotometric measurement of the \halpha~and \nii~fluxes
in a $8\farcs1$-diameter region centered on the nucleus, from
\citep{keel84}. In creating a fluxed \halpha~surface brightness image
we have assumed that the \nii/\halpha~flux ratio is always 0.63,
the value \citet{keel84} found within the nuclear region. 
Any increase, or decrease, in the \nii/\halpha~flux ratio 
will introduce an error into the \halpha~fluxes we quote in this
paper.

Based on observations of edge-on spiral galaxies it is known
that the \nii/\halpha~flux ratio typically increases with height
above the plane. For example, \citet{rand98} finds \nii/\halpha~$\sim 1$ 
-- 1.5 at $z \sim 2$ kpc above the plane of NGC 891, and \citet{tullmann00}
find values of \nii/\halpha~$\sim 1.0\pm{0.1}$ 
at $z \sim 8$ kpc above the plane
of the (possibly starbursting) spiral galaxy NGC 5775.
Unfortunately, there are no published spectroscopic
measurements of the \nii/\halpha~flux ratio high in the halo of NGC 253.
\citet{b-hfq97} measure a \nii/\halpha~flux ratio of $\sim 1$ at large
radii in the plane of NGC 253's disk, beyond the optical edge of the disk,
although this ionized gas may be unrelated to that in the halo of NGC 253.
To give the reader a feeling of the uncertainty this introduces into
the \halpha~fluxes and luminosities we quote later in this paper,
if \nii/\halpha = 1.5 at any point, our estimate, 
using the fixed ratio \nii/\halpha = 0.63, would over-estimate the
true \halpha~flux by 40\%. 

For further details of the narrow-band observations and the associated data
reduction we refer the reader to \citet{hoopes96}.

\subsubsection{X-ray/optical counterparts and astrometric accuracy}
\label{sec:data:astrometry}

We looked for optical counterparts to the {\it Chandra} X-ray sources
in both sets of R-band CCD images.
We used the online catalog of the APM survey \citep{maddox1990} to obtain
coordinates for stars visible in the both sets of CCD images, and hence obtain
astrometric solutions for the CCD images. The rms uncertainty in the
absolute positions in the optical CCD images is $0\farcs15$
(4m images) and $0\farcs20$ (0.6m images).

Comparison between the X-ray positions
and the positions of the 12 optical counterparts found in the S3 
chip field of view, taken from 
the APM catalog, indicates that the absolute coordinates
in the default Chandra astrometry are
accurate to $\sim  0\farcs 7$. 
As greater accuracy is unnecessary for our purposes, we have
not further corrected the {\it Chandra} astrometric solution. All 
X-ray source positions quoted in this paper are based on the
default astrometry.

\section{Spatial morphology of the diffuse X-ray 
emission in the northern halo}
\label{sec:results:spatial_morph}

The soft X-ray ridge or arc seen in earlier {\it ROSAT} observations
of NGC 253 is clearly visible in the {\it Chandra} data  from
energies below $\sim 2$ keV
(see Figs.~\ref{fig:allchips}b \& \ref{fig:halo_diffuse}).
Although there appears to be an excess of point-like sources detected
within the ridge, this is a coincidence. The number of point
sources detected within the halo is consistent with the number expected
based on deeper Chandra surveys of the X-ray background \citep{giaconni2001}. 
The point sources are discussed
in more detail in \S~\ref{sec:halo_point_sources}. 

Based on the {\it Chandra} and {\it ROSAT}
 data it is tempting to interpret the
eastern edge of the ridge as the limb-brightened edge of the wind --
this possibility is discussed in \S~\ref{sec:discussion:models}. 
The spatial resolution of the {\it ROSAT} PSPC
data is not high enough to conclusively prove that the ridge marks a sharp
boundary to the region of X-ray emitting plasma in the east. Unfortunately
the limited angular coverage of the existing {\it Chandra} data does not
extend far enough to the east. {\it XMM-Newton} observations
will better constrain the larger scale morphology of the diffuse
emission, although the
only published description of the 
{\it XMM-Newton} Performance Verification (PV) phase observations
\citep{pietsch2001}
of NGC 253 only discusses  the disk of the galaxy.

The relationship between the diffuse 
X-ray emission and optical \halpha~emission
is explored in more detail in \S~\ref{sec:results:halpha}.

Although the ridge feature dominates the total X-ray flux from the northern
halo, there is diffuse emission away the ridge. The mean X-ray
surface brightness in the north west quadrant of the halo region is
only a factor 2 -- 3 lower than the mean surface brightness within the ridge.
If we interpret the ridge in terms of  limb-brightening, 
the emission that apparently lies within the interior of the wind may  be
from the front and rear walls of the wind.

Previous instruments sensitive to X-rays at energies higher
than 2 keV ({\it ASCA}, {\it BeppoSAX}) did not have sufficient
spatial resolution to determine if there was any diffuse X-ray emission
at these energies (despite some claims to the contrary, \eg \citet{persic98}). 
Within our significantly more sensitive observation of the halo there is
only a marginal $3.3\sigma$ detection of diffuse X-ray flux over the entire
halo region at energies between 2 -- 8 keV.
The ACIS images clearly show (Fig.~\ref{fig:halo_diffuse}d) that
no spatial structure is visible at photon energies greater than
2 keV, once point sources  have been removed from the data.
It is likely that the hard X-ray emission is not related to NGC 253,
but is an artefact of the background subtraction 
(see \S~\ref{sec:results:spectral}).


\begin{figure}[!ht]
\epsscale{1.0}
\plotone{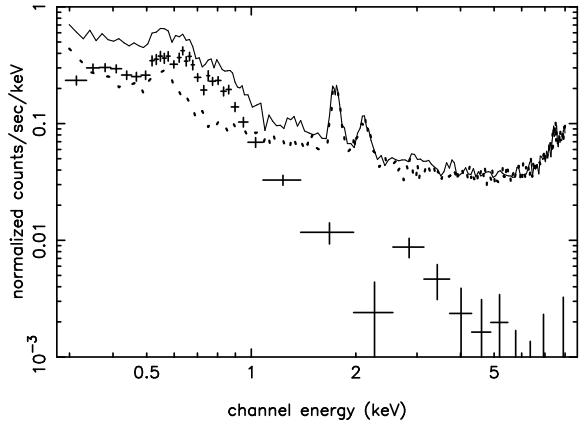}
  \caption{The diffuse emission (\ie 
  point source-subtracted) ACIS-S spectrum of the 
  northern halo region (solid line), predicted background 
  (dotted line) and background-subtracted diffuse spectrum (crosses)
  for the northern halo region.
  Strong instrumental features (e.g at E = 1.8 and 2.0 keV) 
  are well subtracted using the background dataset spectrum. 
  The residual emission in the background-subtracted 
  diffuse spectrum at $E \sim 3 \keV$ is unlikely to
  be a real feature of the X-ray emission from the superwind.
  }
  \label{fig:filtered_spectra}
\end{figure}

\section{Spectral properties of the diffuse X-ray emission}
\label{sec:results:spectral}

We expect thermal X-ray emission from superwinds, from both
physical (see \S~1) and empirical reasons. 
Earlier X-ray spectroscopy of superwinds, including
NGC 253, although best fit by simple thermal models,
gave perplexing results.
These spectral fits appeared to 
imply unphysically low metal abundances (\citet{ptak97}; \citet{tsuru97}),
with iron and $\alpha$-element abundances $\sim 1/20$ and $\sim 1/3$
of the Solar value.
Fitting simple models
to X-ray spectra where there are a range of gas temperatures or
different amounts of absorption can lead to systematically
underestimated metal abundances (see DWH98; \citet{ss98}; \citet{whd}),
which may explain the anomalously low abundances of the earlier
X-ray studies. Accurate abundance determinations are vital for 
obtaining accurate estimates of important plasma properties, as 
the emission integral ($EI = \int n_{\rm e} n_{\rm H} dV$)
trades off directly with the abundance in the
spectral fitting process.

We have therefore expended significant effort in 
constraining the degree of spatial variation in the
spectral properties of the diffuse emission. If it is
possible to spatially resolve, and hence separate,
these spectrally distinct regions to obtain less complex spectra
then more robust estimates of the physical properties of the X-ray
emitting plasma may be obtainable.

Other plausible explanations for low gas phase abundances are
that emission comes from relatively unenriched halo gas that has never been
in the disk of the galaxy, or that sputtering is ineffective at 
removing the refractory elements
from dust grains. Both of these hypotheses can be tested by 
comparing the derived abundances of the gas in the halo with that
from the disk, and comparing abundances of weakly-or-non-depleted elements,
e.g. oxygen, to the abundances of refractory elements such as iron.

We discuss the gross spectral characteristics of the halo diffuse
emission, in comparison the diffuse thermal emission from the
nuclear outflow cone and the diffuse emission within the disk
of NGC 253, in \S~\ref{sec:spectra:compare}.
In \S~\ref{sec:spectra:spatial_variation} we 
investigate whether there is any spatial variation in the spectral
properties of the emission within the halo.
Finding no significant variation on any spatial 
scale, we investigated the spectral
properties of the diffuse emission summed over the entire chip
by fitting various spectral models to the data 
(\S~\ref{sec:spectra:fitting}). As it is easier to predict
a-priori the metal abundance of the X-ray-emitting plasma in
the disk of the galaxy we compare the derived spectral 
properties of the halo to those of the diffuse X-ray emission in the
disk of the galaxy.
Finally, we use the results of this
model fitting to obtain crude estimates of the physical properties
of the X-ray emitting plasma (\S~\ref{sec:spectra:properties}).

\begin{figure*}[!ht]
\epsscale{2.0}
\plotone{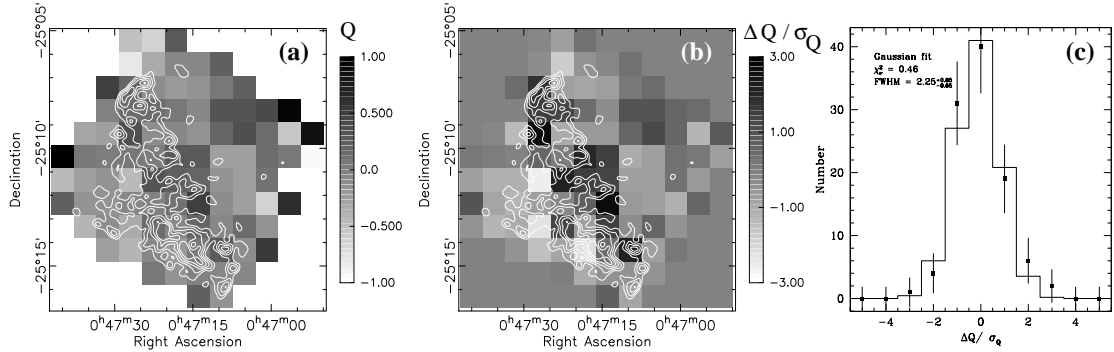}
  \caption{\scriptsize Spatial variation 
  in spectral properties of the diffuse 
  emission in the northern halo: 
  (a) The halo 
  diffuse emission hardness ratio $Q = (H-S)/(H+S)$ as a function of
  position, where the hard energy band $H = 0.6$ -- 1.0 keV and the
  soft energy band $S = 0.3$ -- 0.6 keV. The contours show the 
  $20\arcsec$-FWHM smoothed 0.3 -- 2.0 keV image. The mean hardness
  over the entire chip is $Q_{\rm mean} = -0.035\pm{0.018}$.
  (b) The hardness ratio deviation significance $\Delta Q / \sigma_{Q}$ map.
  (c) The number distribution of hardness ratio deviation significances
  $\Delta Q / \sigma_{Q}$ for the diffuse halo emission,
  fit by a Gaussian model. 
	}
  \label{fig:halo_hardness}
\end{figure*}

\subsection{Spectral characteristics of the halo}
\label{sec:spectra:compare}

The diffuse X-ray spectrum (\ie after emission due to 
point sources and the background has been subtracted)
of the entire northern halo region 
is dominated by emission below photon energies of $1$ keV
(see Fig.~\ref{fig:filtered_spectra}). This is consistent
with the earlier ROSAT PSPC results, which characterized the emission
from the halo as soft thermal emission with $kT \sim 0.2$ keV.
Features attributable to spectrally-unresolved line emission
from highly ionized oxygen, magnesium and possibly iron are
visible in the spectrum, although much weaker than would be expected
for a Solar-metallicity hot plasma. This spectrum looks very similar
to that of the halo region of NGC 4631 (see Fig.~3 in
\citet{wang2001}), another star-forming galaxy that
most probably also hosts a superwind outflow. 
The ACIS spectrum of NGC 253's southern nuclear outflow
cone \citep{strickland2000} is spectrally harder (and therefore
presumably hotter and/or more absorbed) than the diffuse
emission in the halo.

Interestingly, there appears to be very little diffuse emission
at energies above 2 keV. Prior to {\it Chandra} it was
impossible to obtain information on the diffuse emission from
superwinds due to contamination from point source emission
using older, lower spatial resolution, instruments such as {\it ASCA}.

The diffuse spectrum shown in Fig~\ref{fig:filtered_spectra} shows
a peculiar bump in the spectrum at $E \sim 3$ keV, which can
not easily be attributed to either emission from the superwind
itself or explained as an instrumental feature. Instrument-related
features due to silicon in the non-background subtracted spectrum 
($E \sim 2$ keV) subtracted out almost perfectly,
suggesting the 3 keV bump is not a feature present in all ACIS data.
Unresolved line emission from hydrogen-or-helium-like argon
would appear at approximately 3 keV, but to explain this feature
as argon emission from the superwind would require argon abundances
$\ga 20$ times the Solar value, which is highly unlikely given
that we do not see such strong spectral features from the other
$\alpha$-elements.

Emission associated with periods 
of enhanced background in the detector (``flares'') 
can lead to apparent excesses at energies
between 2 -- 4 keV (Markevitch 2000, private communication). We experimented
with more stringent filtering of the data to remove flares, but
both the 2 -- 8 keV count rate and the strength of the 3 keV feature did not
change significantly.  
If a flare is the cause of the observed excess, then such a flare event
must have been extremely long-lived, as we have found same 3 keV feature,
at a similar strength,
present in point-source and background subtracted spectra
from our shorter ACIS observation of NGC 253 taken eleven
days earlier (on 2000 December 16), and in the archived
observations of SN 1999em (ObsID 765, taken 2000 December 16) and
the QSO 3C 220.1 (ObsID 839, taken 2000 December 29).
No such spectral feature is found in the XMM-Newton PV observation
of NGC 253 (Summers 2001, private communication).
We therefor feel confident that  the 3 keV feature is not genuinely
associated with NGC 253. The detection of {\em any} diffuse emission
at energies above 2 keV is of marginal significance in any case
($0.0107\pm{0.0032}$ count s$^{-1}$ in the 2 -- 8 keV energy band), and
is best treated as an upper limit on the hard X-ray emission
from the superwind.

\subsection{Spatial variation in spectral properties}
\label{sec:spectra:spatial_variation}

Prior to attempting to fit the X-ray spectrum of the diffuse
emission, it is important to assess the amount of spatial
variation of the spectral properties within the region
used to obtain the spectrum. We wish to limit the
degree to which any spectrum mixes together emission from
spectrally distinct regions, either of different temperature
or hidden behind different amounts of absorption.

Any spatial variation in spectral properties of the diffuse emission
might also provide important clues as to the origin of the X-ray emission.
For example, if the X-ray emission comes from internal shocks
within the wind, then we might expect increases in characteristic
temperature (or spectral hardness) to directly correlate with
enhancements in X-ray surface brightness. If the X-ray emission
comes from a spatially-resolved constant-pressure conductive interface
(e.g.~as in the \citet{weaver77} wind-blown-bubble model) we
expect the gas temperature to decrease as the density (and surface
brightness) increases.

The limiting factor that controls what range of spatial scales
over which spectral variation can be investigated 
in such an analysis using {\it Chandra} is signal-to-noise per spatial
bin, rather
than the spatial resolution of the instrument. With arcsecond spatial
resolution there is no effective ``cross-talk'' between spatial
bins, and point sources can easily be removed without leading to
significant loss of area. Given the surface brightness of the halo diffuse
emission and the instrumental and X-ray backgrounds, we found
that obtaining hardness ratios with any meaningful 
statistical certainty necessitates using spatial bins at least
$30\arcsec$-wide. This corresponds to physical scales of $\gtrsim 380 \pc$
at the distance of NGC 253.

\begin{deluxetable}{lcccc}
 \tabletypesize{\scriptsize}%
\tablecolumns{5} 
\tablewidth{0pc} 
\tablecaption{Spectral hardness ratios in the halo, disk and nuclear
	outflow cone
	\label{tab:regions_hardness}}
\tablehead{ 
\colhead{Region} & \colhead{Count rate}
	& \colhead{$Q_{A}$} & \colhead{$Q_{B}$} & \colhead{$Q_{C}$} \\
\colhead{(1)} 
	& \colhead{(2)}
	& \colhead{(3)} & \colhead{(4)} & \colhead{(5)}
	}
\startdata
Northern halo    & $0.2092\pm{0.0053}$ 
	& $-0.035\pm{0.018}$ & $-0.755\pm{0.025}$ & $-0.909\pm{0.034}$ \\
NE halo & $0.0521\pm{0.0025}$ 
	& $+0.002\pm{0.037}$ & $-0.799\pm{0.051}$ & $-0.795\pm{0.061}$ \\
NW halo & $0.0292\pm{0.0023}$ 
	& $-0.018\pm{0.061}$ & $-0.633\pm{0.073}$ & $-0.827\pm{0.103}$ \\
SE halo & $0.0521\pm{0.0025}$ 
	& $-0.103\pm{0.034}$ & $-0.783\pm{0.045}$ & $-1.000\pm{0.064}$ \\
SW halo & $0.0750\pm{0.0026}$ 
	& $-0.018\pm{0.029}$ & $-0.752\pm{0.037}$ & $-0.888\pm{0.047}$ \\
Disk    & $0.4527\pm{0.0073}$ 
	& $0.188\pm{0.017}$ & $-0.615\pm{0.018}$ & $-0.832\pm{0.021}$ \\
Cone (east limb) & $0.0184\pm{0.0020}$ 
	& $0.139\pm{0.074}$ & $-0.666\pm{0.088}$ & $-0.992\pm{0.163}$ \\
Cone (center) & $0.0084\pm{0.0019}$ 
	& $0.284\pm{0.117}$ & $-0.689\pm{0.154}$ & $-0.994\pm{0.329}$ \\
Cone (west limb) & $0.0423\pm{0.0024}$ 
	& $0.404\pm{0.053}$ & $-0.545\pm{0.051}$ & $-0.979\pm{0.082}$ \\
\enddata 
\tablecomments{Errors are quoted at 68\% percent confidence.
	Column (1): The halo and disk regions are defined
        in Table~\ref{tab:halo_region}. Cone refers to the clearly
        limb-brightened region of the southern nuclear outflow cone,
        as defined in \citet{strickland2000} 
        (2): Background and source-subtracted
	ACIS-S3 count rate (units of counts/s)
	in the 0.3 -- 8.0 keV energy band. These count rates 
	have {\em not} been
        corrected for the area lost due to point source removal.
	(3): Hardness ratio Q=(H-S)/(H+S), where H is the count rate in the
        0.6 -- 1.0 keV energy band and S is the count rate in the
        0.3 -- 0.6 keV energy band.
	(4): Hardness ratio using count rates in
        the 1.0 -- 2.0 keV (H) and 0.3 -- 1.0 keV (S)
	energy bands.
	(5): Hardness ratio using count rates in 
        the 2.0 -- 8.0 keV (H) and 0.3 -- 2.0 keV (S)
	energy bands.
	}
\end{deluxetable}

\subsubsection{Hardness ratio maps of the halo}

Fig.~\ref{fig:halo_hardness}a shows a hardness map of the diffuse
emission in the northern halo, constructed using $1\arcmin$-wide
pixels. The hardness ratio is defined
as $Q = (H-S)/(H+S)$, where $H$ is the count rate in hard
band and $S$ is the count rate in the soft band.
Defined in this way, the hardness
ratio  always lies between -1 (soft) and 1 (hard).

We use data from the 0.6--1.0 keV energy band as the hard band,
and the 0.3--0.6 keV energy band as the soft band, as the observed halo
region spectrum has roughly equal number of counts in these two bands
(hence maximizing the S/N in each band).
A harder (higher temperature and/or more absorbed) 
X-ray spectrum in any region gives a positive hardness
ratio and a soft spectrum a negative hardness ratio. 

For the low X-ray count rates found in superwinds, hardness ratio
maps do not give a statistically robust estimate of variations
in hardness due to Poisson fluctuations. 
Fig.~\ref{fig:halo_hardness}b is a hardness ratio
deviation significance map, which maps the statistical significance
of any deviation from the mean hardness of the diffuse emission.
For each pixel $i,j$ in the image, the hardness ratio deviation significance
is $(Q_{i,j}-Q_{\rm mean})/\sigma_{Q_{i,j}}$, where $Q_{i,j}$ is the 
hardness ratio at that point, $\sigma_{Q_{i,j}}$ is the uncertainty in
the hardness ratio at that point, and $Q_{\rm mean}$ is the mean hardness
of the diffuse emission. The procedure used to construct the
hardness ratio and hardness ratio deviation significance maps is
discussed in Appendix~\ref{sec:app:hardness}.
Note that the calibration problems discussed in \S~\ref{sec:data:calibration}
have negligible effect on these hardness ratio studies,
given the broad energy bands used.

We find only marginally-significant deviations ($\lesssim 3\sigma$) 
from the mean hardness ratio in the halo. These do not appear to correlate with
the structure seen in the soft X-ray surface brightness map, which
suggests that these hardness variations are noise. Support for this
hypothesis comes from the number distribution of hardness ratio
deviation significances, which is well fit with a model of Gaussian
random noise (Fig.~\ref{fig:halo_hardness}c). Consistent results
are obtained if larger or smaller bin sizes ($2\arcmin$ or $30\arcsec$-wide
pixels) are used for the hardness ratio
maps.

On angular scales of $30\arcsec$ -- $2\arcmin$ ($\sim 380$ to 1500 pc)
we conclude that there is no
statistically significant variation in the spectral properties of the
diffuse emission from the superwind.

\begin{figure*}[!t]
\epsscale{2.0}
\plotone{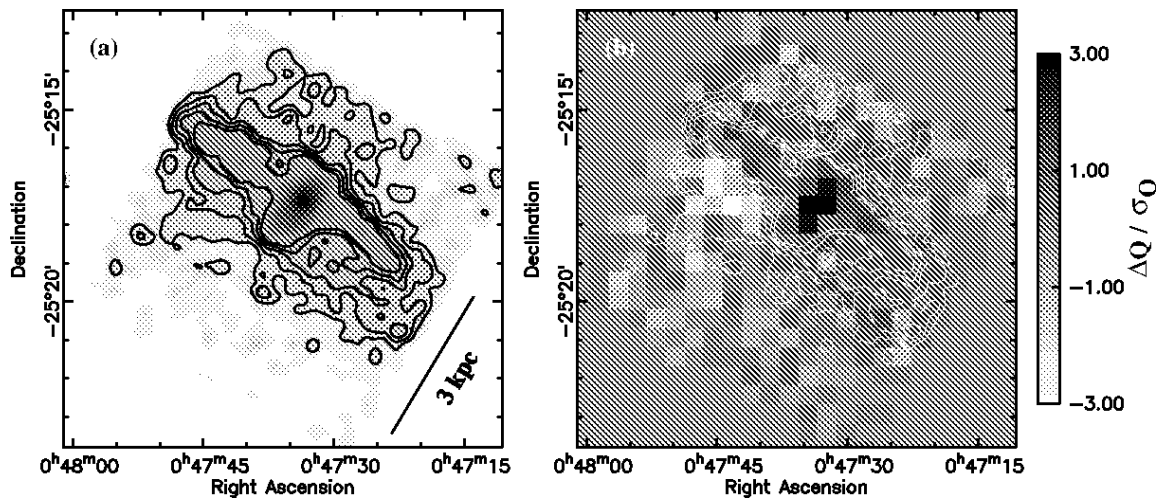}
\caption{(a) Chandra ACIS-S2 0.3 -- 2.0 keV image of the diffuse X-ray
  emission from the disk of NGC 253, along with the inner halo 
  to the SE and NW of the disk, smoothed with a $20\arcsec$ FWHM Gaussian
  mask. Point sources and background emission have been subtracted 
  from the data. Contours are shown at signal-to-noise ratios of 2, 3, 4,
  5 \& 6 above the background within a $20\arcsec$-diameter aperture.
  The grey scale image is shown on a logarithmic intensity scale between
  $10^{-7}$ to $3\times 10^{-4}$ counts s$^{-1}$ arcsec$^{-2}$.
  (b) The diffuse emission hardness ratio deviation significance
  map. White contours correspond to those shown in panel (a). The only
  regions that deviate significantly from the mean spectral hardness are
  the nuclear outflow cone (spectrally harder than the disk) 
  and a small region near the base of the south eastern plume 
  of the superwind (spectrally softer than the rest of the diffuse
  X-ray emission from the disk). The pixel scale is $30\arcsec$.}
\label{fig:disk_hardness}
\end{figure*}

\subsubsection{Hardness ratio maps of the disk}
\label{sec:disk_hardness}
We applied the same hardness ratio mapping technique to investigate
the degree of spectral variation in the diffuse X-ray emission from
the disk of the galaxy.

We used the data from the 43 ks observation in which the disk of the
galaxy was placed on the front-illuminated S2 chip. This provides
approximately twice the total number of counts as the shorter
observation with the disk in the S3 chip, and has the added advantage
that the spatial coverage of the S2 chip extends a few kpc into the
southern halo, partially covering the south eastern spur of the
superwind seen in Fig.~\ref{fig:pspc_chips}.

As with the hardness maps of the halo, we used 0.3 -- 0.6 keV
and 0.6 -- 1.0 keV energy bands, having subtracted both point sources
and background from the data. A point source-subtracted image of the
diffuse emission is shown along with the hardness
ratio deviation significance map in Fig.~\ref{fig:disk_hardness}.

Only two regions show hardness
ratios that deviate significantly from the mean spectral hardness
of the diffuse disk emission. Diffuse emission from the central $\sim 1 $
kpc (comprising the obscured
starburst region, the southern nuclear outflow cone 
and the heavily absorbed northern
outflow cone) appears systematically harder than the rest of the disk.
Note that this region had been excluded when initially calculating the
mean spectral hardness of the disk.
To the east of the nucleus, somewhat away from the brightest emission
from the disk, is a $\sim 3\arcmin \times 1.5\arcmin$ region that
is significantly softer than the average disk emission.
\citet{pietsch2001} noted that this region [which they call NE(S)]
appeared spectrally soft
in the {\it XMM-Newton} EPIC PN data, and suggested it came from less
absorbed gas above the disk of the galaxy. In the {\it Chandra} data
this region seems to lie at the base of the base of the south
eastern plume of H$\alpha$ and soft X-ray emission (compare
Fig.~\ref{fig:disk_hardness} with 
Figs.~\ref{fig:pspc_chips} \& \ref{fig:halpha_images}), possibly indicating
an association with the superwind.

\begin{figure*}[!ht]
\epsscale{2.0}
\plotone{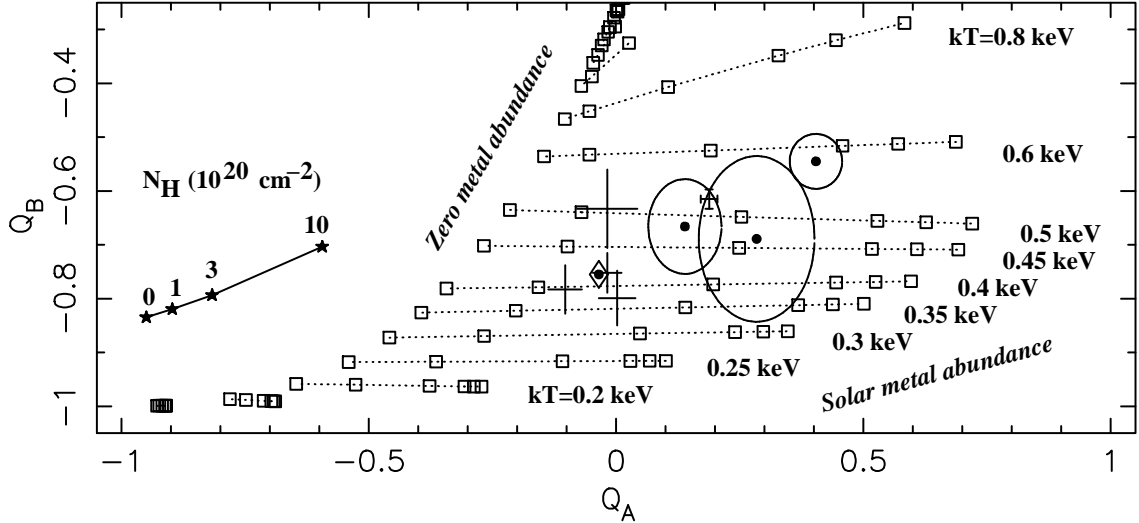}
\caption{A hardness ratio $Q_{a}$ vs. hardness ratio $Q_{B}$ plot, showing
  the observed hardness ratios of the northern halo diffuse emission 
  (diamond), the four separate quadrants of the northern halo (crosses), the
  diffuse disk emission (barred cross) 
  and the center, east and west limbs of
  the southern nuclear outflow cone (ellipses). 
  The hardness ratios $Q_{A}$ and 
  $Q_{B}$ are defined in Table.~\ref{tab:regions_hardness}. Theoretical
  hardness ratios are shown for an {\em unabsorbed} 
  single phase hot plasma with 
  $kT$ = 0.1, 0.15, 0.2, 0.25, 0.3, 0.35, 0.4, 0.45, 0.5, 0.6, 0.8, 1.0 and
  1.25 keV, $10 \times 10^{20} \pcmsq$, where the open squares represent
  $Z = 0, 0.02, 0.1, 0.3, 0.5$ and $1 \times Z_{\sun}$ moving from
  left to right. The starred points on the left of the plot show the
  degree to which these hardness ratios change depending on the foreground
  hydrogen column density. 
  The relative softness of all the diffuse emission regions in ratio $Q_{A}$
  forces the fits to give low  metal abundances
  when using single phase spectral models. Error bars are $1\sigma$ 
	confidence levels.}
\label{fig:color_color_plot}
\end{figure*}

\subsubsection{Spectral variation over larger spatial scales}

We searched for spectral variation in the halo diffuse emission
between the $4.2\arcmin$-square
quadrants that make up the northern halo region.
These quadrants cover a range in diffuse X-ray
surface brightness, from the faint NW region away from the X-ray arc,
through the moderately brighter SE and NE regions to the brightest
diffuse emission in the SW region. If there is any significant
variation in the spectral properties of the diffuse emission
over $\sim 3$ kpc scales, then this should be apparent
comparing the diffuse emission spectra from these four regions.

We found only weak differences between the regions in any of the
hardness ratios we considered (see Table~\ref{tab:regions_hardness}),
none of which are statistically significant at more than the
$2\sigma$ level. Based on Fig.~\ref{fig:color_color_plot} this
limits variations in characteristic temperature in the halo to $\la 50 \%$,
or changes in hydrogen column of $\la 0.5$ dex.

\subsection{Spectral fitting}
\label{sec:spectra:fitting}

Given the lack of significant spatial variation in the spectral properties
of the diffuse X-ray emission over the halo region, we have chosen
to fit spectral models to the ACIS spectrum of the entire halo region.

The count rate in the diffuse spectrum
at energies above 2 keV is only marginally significant, 
and as discussed previously the detected counts
are likely to be due to low-level systematic problems 
with the background subtraction. For this reason we choose to
fit the data only in the energy range 0.3 -- 2.5 keV to avoid biasing 
the fits. Experiments of fitting the data using a wider energy range
give similar and consistent results, in several cases with lower
values of reduced $\chi^{2}$ due to the addition of 
data points with very large uncertainties.  Within the 0.3 -- 2.5 keV
energy range the background and source subtracted count rate from
the diffuse halo emission is $0.2014\pm{0.0037}$ counts per second 
(a total of $7973\pm{146}$ counts over 39588 seconds).

\subsubsection{Spectral models}
 
We considered a variety of models that might reasonably be expected
to represent the thermal X-ray emission from the hot gas in superwinds.
The standard hot plasma models used to fit the soft thermal component
in superwinds assume collisional ionization equilibrium (CIE). We have
used four of the currently available CIE hot plasma codes: {\sc Mekal} 
\citep{mekal}, {\sc apec} \citep{apec}, 
the Raymond-Smith code \citep{raymond-smith} and the 
{\sc Equil} \citep{borkowski2001} code. 
When combined with a photoelectric absorption model \citep{morrison83}
a hot plasma model is parameterized in terms of equivalent hydrogen column
$\nH$, gas thermal temperature $kT$, metal abundances relative to 
Solar\footnote{Note that abundances within {\sc Xspec} are based 
on \citet{anders89}, which over-estimates the
Fe abundance by 0.17 dex compared to more recent estimates 
(\eg \citet{grevesse93}).} and an overall normalization which depends on
the emission integral $EI = \int n_{\rm e} n_{\rm H} dV$.
Non-ionization equilibrium (NIE) models, using the code
developed by \citet{borkowski2001}, are further characterized by
an ionization timescale defined as the product of the
electron density and the age $n_{\rm e} \tau$.
The {\sc Nei} model represents initially cold neutral gas which has been
suddenly heated, in which the ionization state of the ions is lower
than in a CIE plasma of the same kinetic (electron) 
temperature. Other non-ionization equilibrium models with
different ionization histories (\eg cooling gas more ionized than
a same-temperature CIE plasma), such the {\sc Gnei} and {\sc Pshock}
models in {\sc Xspec}, give similar or worse fits to the spectra
than the {\sc Nei} model does. For the sake of brevity we will
not discuss these other NIE models again. Higher resolution
spectroscopy will be needed to robustly constrain the ionization
state and history of the X-ray emitting gas in superwinds. Our primary
interest in the results of the NIE models is whether ionization effects
significantly alter the fitted metal abundances.

These spectra only constrain the abundances of
those elements that produce strong line
emission in the {\it Chandra} energy band, and which produce lines
or line complexes at energies not covered by emission from other
elements that produce stronger line emission. We have calculated
the fraction of emission due to lines of each element for a range
of plasma temperatures using the {\sc Mekal} code. 
Oxygen and iron abundances are likely to be the best-constrained
in ACIS spectra of soft thermal X-ray emission due to
their intensity\footnote{For a plasma with Solar element abundances
oxygen (iron) provides 58\% (14\%) of the ACIS-S 0.3 -- 2.5 keV 
energy band count rate
from a $kT = 0.2$ keV, $\nH = 3 \times 10^{20} \pcmsq$ plasma, and
13\% (61\%) of the emission from a $kT = 0.5$ keV plasma with the
same hydrogen column.},
followed by magnesium and silicon abundances due to their clear
separation in energy space from other spectral features.
In principle element abundances of
C, N, Ne \& S can also be constrained, although the low
energy of the main C, N and S lines, and for Ne confusion with the strong
emission from Fe, makes abundances determinations more difficult.
Confusion with Fe is a major problem for determining Ni abundances.
We can not realistically constrain Na, Al, Ar \& Ca abundances given the
low emissivities and line confusion. For the purposes of the spectral
fitting we have fixed the relative abundances of Na, Al, Ca and Ni
to be equal to that of Fe (\ie [X/H] = [Fe/H] where X is the element
in question),  as these are all refractory elements and depletion
onto dust grains may be significant in the X-ray emitting plasma.
The Ar relative abundance was made equal to the Ne abundance. 

\subsubsection{Single phase spectral models}
\label{sec:spectra:single}

Single temperature spectral models of the type discussed in the
previous section all provided statistically
unacceptable fits to the halo spectrum. 
Reduced chi-squared values
in the range $\chi^{2}_{\nu} = 1.22$ to 1.99 were obtained
for between 77 and 85 degrees of freedom, depending on which
model was used.

Fitted temperatures where typically $kT \sim 0.4$ keV, the extremes
being 0.25 and 0.69 keV. Best fit metal abundances were always extremely
low, typically $Z = 0.02\pm{0.02} Z_{\sun}$.
Fitted oxygen and iron abundances were $Z_{O}
\sim 0.05 Z_{O, \sun}$ (typical 90\% uncertainty $\la 0.06 Z_{\sun}$)
and $Z_{Fe} \sim 0.03 Z_{Fe, \sun}$ (typical 
90\% uncertainty $\la 0.07 Z_{\sun}$).
Note that this is very similar to the
temperature and abundances  one would infer from a simple
hardness ratio analysis (see Fig.~\ref{fig:color_color_plot}).
The simple explanation is that the fitting process must match the
broad band spectral shape of the halo spectrum. Direct diagnostics
of the temperature of the plasma  (\eg using flux ratios of 
the hydrogen to helium-like oxygen line complexes), or
of the metal abundance (\eg the apparent equivalent width of 
line emission above the continuum) play only a secondary role in
constraining the fitted parameters. Metal line equivalent
widths are extremely difficult to measure from ACIS resolution
spectra directly, as the continuum is buried under the merged
blend of line emission\footnote{Only $\sim 5$ to 10\% of the counts from a
Solar abundance plasma are from the continuum for gas
temperatures $0.1 \lesssim kT \lesssim 1$ keV, and line emission
covers the continuum at almost all energies below 2 keV.}.

\begin{deluxetable}{lccccccccccc}
\tabletypesize{\scriptsize}
\rotate
\tablecolumns{12} 
\tablewidth{0pc} 
\tablecaption{Multi-component spectral model fits 
	to the diffuse halo and disk emission
	\label{tab:fits:multi_solar}}
\tablehead{ 
\colhead{Model} & \colhead{$\nH$} 
	& \colhead{$kT_{1}$ or $\Gamma$} & \colhead{$kT_{2}$}
	& \colhead{$\log (n_{\rm e} \tau)_{1}$} 
	& \colhead{$\log (n_{\rm e} \tau)_{2}$}
	& \colhead{$\alpha_{\rm DEM}$} & \colhead{$FWHM$}
	& \colhead{$K_{1}$} & \colhead{$K_{2}$} 
 	& \colhead{$\Delta E$}
	& \colhead{$\chi^{2}_{\nu}$ ($\chi^{2}$/$\nu$)} \\
\colhead{(1)} 
	& \colhead{(2)}
	& \colhead{(3)} & \colhead{(4)} & \colhead{(5)}
	& \colhead{(6)} & \colhead{(7)} & \colhead{(8)}
	& \colhead{(9)} & \colhead{(10)}
	& \colhead{(11)} & \colhead{(12)}
	}
\startdata
\cutinhead{Northern halo spectral fits}
2Tvmekal & $6.1^{+8.0}_{-1.5}$ 
	& $0.24^{+0.07}_{-0.04}$ & $>0.4$\tablenotemark{a}  
	 & \nodata & \nodata & \nodata 
	& \nodata 
	& $34.16^{+88.79}_{-25.37}$ & $1.66^{+1.57}_{-1.25}$
	& $-38^{+21}_{-24}$ & 1.19 (91.8/77) \\
2Tvapec & $5.3^{+8.3}_{-2.1}$ 
	& $0.24^{+0.03}_{-0.04}$ & $0.71^{+3.18}_{-0.36}$  
	 & \nodata & \nodata & \nodata 
	& \nodata 
	& $40.96^{+95.12}_{-30.47}$ & $4.79^{+11.39}_{-4.33}$
	& $-27^{+25}_{-16}$ & 1.11 (85.3/77) \\
2Tvnei & $5.0^{+17.4}_{-3.0}$ 
	& $0.33^{+0.18}_{-0.33}$ & $14.2$\tablenotemark{b}  
	 & $11.17^{+5.83}_{-0.65}$ & \nodata & \nodata 
	& \nodata 
	& $21.53^{+17.80}_{-11.83}$ & $0.87^{+1.99}_{-0.87}$
	& $-34^{+25}_{-19}$ & 1.23 (93.7/76) \\
2Ivnei & $2.1^{+8.2}_{-1.8}$ 
	& $0.70^{+0.65}_{-0.35}$ & \nodata  
	 & $9.88^{+1.81}_{-9.88}$ & $10.99^{+0.61}_{-0.49}$ & \nodata 
	& \nodata 
	& $1.06^{+2.02}_{-1.06}$ & $4.23^{+13.48}_{-3.20}$
	& $-29^{+25}_{-15}$ & 1.13 (86.1/76) \\
GTvmekal & $5.5^{+7.0}_{-2.8}$ 
	& $0.26^{+0.10}_{-0.19}$ & \nodata
	 & \nodata & \nodata & \nodata
	& $0.446^{+1.386}_{-0.445}$
	& $31.61^{+63.18}_{-26.21}$ & \nodata
	& $-32^{+24}_{-22}$ & 1.29 (100.4/78) \\
PLTvmekal & $5.7^{+4.5}_{-2.8}$ 
	& $0.20^{+0.09}_{-0.04}$ & $>18.1$\tablenotemark{a}  
	 & \nodata & \nodata & $-2.42^{+2.28}_{-7.40}$ 
	& \nodata 
	& $31.21^{+31.46}_{-20.71}$ & \nodata
	& $-33^{+22}_{-25}$ & 1.25 (96.1/77) \\
PL+vapec & $6.4^{+3.4}_{-3.6}$ 
	& $3.3^{+0.5}_{-2.9}$  & $0.24^{+0.04}_{-0.03}$
	 & \nodata & \nodata & \nodata 
	& \nodata 
	& $7.83^{+17.20}_{-2.72}$ & $3.72^{+248.2}_{-3.17}$
	& $-28^{+28}_{-20}$ & 1.16 (89.0/77) \\
\cutinhead{Disk diffuse emission spectral fits}
2Tvmekal & $4.4^{+2.7}_{-2.1}$ 
	& $0.17^{+0.12}_{-0.10}$ & $0.57^{+0.06}_{-0.08}$
	 & \nodata & \nodata & \nodata 
	& \nodata 
	& $5.98^{+40.27}_{-4.41}$ & $21.21^{+1.36}_{-9.72}$
	& $-2^{+21}_{-10}$ & 1.39 (128.3/92) \\
2Tvapec & $4.7^{+4.2}_{-2.4}$ 
	& $0.17^{+0.07}_{-0.11}$ & $0.56^{+0.13}_{-0.06}$
	 & \nodata & \nodata & \nodata 
	& \nodata 
	& $8.25^{+55.18}_{-3.20}$ & $32.37^{+22.53}_{-18.47}$
	& $-1^{+21}_{-7}$ & 1.34 (123.6/92) \\
2Ivnei & $2.8^{+4.1}_{-2.4}$ 
	& $1.24^{+2.14}_{-0.74}$& \nodata
	 & $>9.60$\tablenotemark{a} &  $>10.10$\tablenotemark{a} 
	& \nodata 
	& \nodata 
	& $4.65^{+17.01}_{-3.07}$ & $3.97^{+19.74}_{-3.32}$
	& $-5^{+15}_{-17}$ & 1.36 (124.1/91) \\
2Tvnei & $2.9^{+3.3}_{-1.8}$ 
	& $0.49^{+0.24}_{-0.16}$ & $>0.60$\tablenotemark{a}
	 & $10.54^{+0.29}_{-0.11}$ & \nodata & \nodata 
	& \nodata 
	& $11.16^{+9.24}_{-5.96}$ & $5.39^{+2.59}_{-3.34}$
	& $-5^{+16}_{-16}$ & 1.21 (109.9/91) \\
GTvmekal & $5.2^{+3.3}_{-2.8}$ 
	& $0.30^{+0.21}_{-0.20}$ & \nodata
	 & \nodata & \nodata & \nodata 
	& $0.76^{+0.56}_{-0.34}$
	& $41.31^{+201.81}_{-28.81}$ & \nodata
	& $-3^{+11}_{-12}$ & 1.44 (133.8/93) \\
PLTvmekal & $5.0^{+3.6}_{-2.6}$ 
	& $0.08^{+0.17}_{-0.08}$ & $0.82^{+2.82}_{-0.28}$
	 & \nodata & \nodata & $0.51^{+1.16}_{-2.48}$
	& \nodata 
	& $36.95^{+50.97}_{-19.67}$ & \nodata
	& $-3^{+22}_{-12}$ & 1.41 (129.9/92) \\
PL+vapec & $6.1^{+4.0}_{-4.1}$ 
	& $2.1^{+0.8}_{-1.6}$  & $0.25^{+0.11}_{-0.05}$
	 & \nodata & \nodata & \nodata 
	& \nodata 
	& $17.05^{+9.66}_{-10.15}$ & $3.71^{+73.12}_{-3.64}$
	& $-5^{+18}_{-10}$ & 1.38 (126.8/92) \\
\enddata 
\tablecomments{All errors are 90\% confidence for a number of interesting
	parameters equal to the number of free parameters in the model
	(13, 14 or 15 free parameters depending on the model).
	Best-fit element abundances are given in 
	Table~\ref{tab:fits:metals:multi}. 
	Column (1): {\sc Xspec} model name. Variable metal abundance
        hot plasma codes that 
	assume collisional ionization equilibrium are {\tt vmekal} 
	\& {\tt vapec}, while {\tt vnei} is a variable abundance
        non-ionization equilibrium plasma model.
        A prefix of 2T or 2I denotes that the model is a two temperature
        or two ionization timescale model, while GT and PLT refer to
        power law or Gaussian distributions of emission integral
        as a function of temperature. The {\tt PL+vapec} model is
	a two component power law plus {\tt vapec} thermal plasma model. 
        See text for details.
	Column (2): Hydrogen column density in units of $10^{20} \pcmsq$.
	Columns (3) \& (4): Gas temperatures in keV. For the 2T models
        column (3) is the temperature of the first component, and column (4)
        the temperature of the second component. For the PLT model (3) is
        the minimum temperature and (4) the maximum temperature over which 
        the power law in emission integral holds. For the GT model
        (3) is the central temperature of the Gaussian emission
        integral distribution. For the {\tt PL+vapec} model column (3)
        is the photon index of the power law $\Gamma$, where the
        flux $f(E) \propto E^{-\Gamma}$.
	Columns (5) \& (6): Logarithm of the ionization timescale (cgs units)
	in the 2Tvnei and 2Ivnei models.
	Column (7): The slope $\alpha$ of the emission integral distribution in
        the PLT model, where $EI(T) \propto T^{\alpha}$.
	Column (8): Full width at half maximum (dex) of the Gaussian
	distribution in X-ray emission integral used
	in the GT model. Note that the distribution
        is Gaussian in $\log T$.
	Columns (9) \& (10): Model component normalizations, where 
	$K = 10^{-10} EI / 4 \pi D^{2}$.
	$D$ is the distance to the source, and the emission integral
	$EI = \int n_{\rm e} n_{\rm H} dV$ is the volume integral of the 
	electron and hydrogen number densities. For the {\tt PL+vapec} 
	model column (9) is the power law normalization in units
	of $10^{5}$ photons s$^{-1}$ keV$^{-1}$ cm$^{-2}$ at $E=1$ keV.
	Column (11): Best-fit energy shift 
        applied to the spectral response in eV.
	Column (12): Fit statistics: the reduced chi-squared 
	($\chi^{2}_{\nu}$), the value of $\chi^{2}$ and the number of degrees
	of freedom. 
	}
\tablenotetext{a}{At 90\% confidence only a lower limit on the value of
	this parameter could be obtained.}
\tablenotetext{b}{At 90\% confidence this parameter is unconstrained.}
\tablenotetext{c}{Parameter fixed at this value during fitting.}
\end{deluxetable}

\begin{deluxetable}{lcccccccc}
 \tabletypesize{\scriptsize}%
\tablecolumns{9} 
\tablewidth{0pc} 
\tablecaption{Metal abundances from the multi-component spectral models
	\label{tab:fits:metals:multi}}
\tablehead{ 
\colhead{Model} & \colhead{C} & \colhead{N} 
	& \colhead{O} & \colhead{Ne, Ar} 
	& \colhead{Mg} & \colhead{Si} 
	& \colhead{S} & \colhead{Fe, Na, Al, Ca} \\
\colhead{(1)} 
	& \colhead{(2)}
	& \colhead{(3)} & \colhead{(4)} & \colhead{(5)}
	& \colhead{(6)} & \colhead{(7)} & \colhead{(8)}
	& \colhead{(9)}
	}
\startdata
\cutinhead{Northern halo spectral fits}
2Tvmekal & $0.168^{+1.357}_{-0.168}$ & $0.000^{+0.756}_{-0.000}$
	& $0.042^{+0.109}_{-0.021}$ & $0.097^{+0.178}_{-0.067}$ 
	& $0.306^{+0.965}_{-0.306}$ & $0.204^{+3.242}_{-0.204}$ 
	& $0.000^{+0.316}_{-0.000}$ & $0.053^{+0.153}_{-0.032}$  \\
2Tvapec & $0.226^{+1.634}_{-0.0226}$ & $0.000^{+0.689}_{-0.000}$
	& $0.038^{+0.319}_{-0.023}$ & $0.037^{+0.147}_{-0.037}$ 
	& $0.226^{+2.086}_{-0.226}$ & $0.252^{+1.718}_{-0.252}$  
	& $0.000^{+0.440}_{-0.000}$ & $0.050^{+0.217}_{-0.032}$ \\
2Tvnei & $0.397^{+4.262}_{-0.397}$ & $0.000^{+0.581}_{-0.000}$
	& $0.026^{+0.227}_{-0.015}$ & $0.056$\tablenotemark{a} 
	& $0.188^{+2.685}_{-0.188}$ & $0.444^{+8.161}_{-0.444}$ 
	& $0.000^{+1.209}_{-0.000}$ & $0.037^{+0.115}_{-0.024}$  \\
2Ivnei & $0.212^{+2.443}_{-0.212}$ & $0.33^{+19.34}_{-0.33}$
	& $0.092^{+0.825}_{-0.060}$ & $0.026^{+0.445}_{-0.026}$ 
	& $0.163^{+0.359}_{-0.163}$ & $0.125^{+2.098}_{-0.125}$ 
	& $0.000^{+0.000}_{-7.370}$ & $0.107^{+0.294}_{-0.107}$  \\
GTvmekal & $0.190^{+7.043}_{-0.190}$ & $0.000^{+2.185}_{-0.000}$
	& $0.064^{+0.475}_{-0.032}$ & $0.064^{+0.165}_{-0.064}$ 
	& $0.179^{+0.500}_{-0.179}$ & $0.227^{+1.119}_{-0.227}$ 
	& $0.000^{+0.670}_{-0.000}$ & $0.032^{+0.221}_{-0.017}$  \\
PLTvmekal & $0.228^{+1.178}_{-0.228}$ & $0.000^{+0.536}_{-0.000}$
	& $0.057^{+0.114}_{-0.030}$ & $0.075^{+0.179}_{-0.075}$ 
	& $0.215^{+0.332}_{-0.215}$ & $0.299^{+2.036}_{-0.299}$ 
	& $0.000^{+0.446}_{-0.000}$ & $0.043^{+0.110}_{-0.025}$  \\
PL+vapec & $2.2^{+16.3}_{-2.2}$ & $0.0^{+4.2}_{-0.0}$
	& $0.45^{+0.35}_{-0.22}$ & $0.85^{+0.94}_{-0.66}$ 
	& $3.3$\tablenotemark{a} & $4.3$\tablenotemark{a} 
	& $0.0^{+11.7}_{-0.0}$ & $0.81^{+0.45}_{-0.69}$  \\
\cutinhead{Disk diffuse emission spectral fits}
2Tvmekal & $0.000^{+1.437}_{-0.000}$ & $0.019^{+1.076}_{-0.019}$
	& $0.212^{+0.476}_{-0.152}$ & $0.045^{+0.180}_{-0.045}$ 
	& $0.150^{+0.476}_{-0.150}$ & $0.340^{+0.618}_{-0.170}$ 
	& $1.072^{+2.454}_{-1.072}$ & $0.072^{+0.063}_{-0.035}$  \\
2Tvapec & $0.000^{+1.033}_{-0.000}$ & $0.304^{+1.923}_{-0.304}$
	& $0.152^{+0.463}_{-0.128}$ & $0.063^{+0.292}_{-0.063}$ 
	& $0.096^{+0.471}_{-0.096}$ & $0.275^{+0.978}_{-0.263}$ 
	& $1.061^{+3.929}_{-1.061}$ & $0.049^{+0.074}_{-0.025}$  \\
2Ivnei & $0.000^{+0.085}_{-0.000}$ & $0.127^{+0.160}_{-0.000}$
	& $0.042^{+0.059}_{-0.023}$ & $0.019^{+0.071}_{-0.019}$ 
	& $0.105^{+0.187}_{-0.105}$ & $0.264^{+0.710}_{-0.264}$ 
	& $0.686^{+0.713}_{-0.686}$ & $0.141^{+0.222}_{-0.098}$  \\
2Tvnei & $0.019^{+0.352}_{-0.019}$ & $0.071^{+0.202}_{-0.070}$
	& $0.032^{+0.028}_{-0.016}$ & $0.037^{+0.094}_{-0.037}$ 
	& $0.097^{+0.260}_{-0.097}$ & $0.212^{+0.411}_{-0.212}$ 
	& $0.239^{+0.761}_{-0.239}$ & $0.116^{+0.201}_{-0.071}$  \\
GTvmekal & $0.000^{+1.170}_{-0.000}$ & $0.021^{+1.367}_{-0.021}$
	& $0.102^{+0.340}_{-0.048}$ & $0.061^{+0.184}_{-0.061}$ 
	& $0.194^{+0.574}_{-0.194}$ & $0.395^{+0.986}_{-0.395}$ 
	& $0.410^{+1.009}_{-0.410}$ & $0.088^{+0.202}_{-0.039}$  \\
PLTvmekal & $0.000^{+0.587}_{-0.000}$ & $0.000^{+1.071}_{-0.000}$
	& $0.126^{+0.239}_{-0.082}$ & $0.017^{+0.182}_{-0.017}$ 
	& $0.180^{+0.503}_{-0.180}$ & $0.361^{+0.690}_{-0.350}$ 
	& $0.601^{+1.046}_{-0.601}$ & $0.081^{+0.078}_{-0.039}$  \\
PL-vapec & $0.0^{+18.8}_{-0.0}$ & $2.6^{+39.6}_{-2.6}$
	& $0.73^{+9.52}_{-0.64}$ & $2.4^{+18.7}_{-2.0}$ 
	& $4.4$\tablenotemark{a} & $14.1$\tablenotemark{a}
	& $4.2^{+34.8}_{-4.2}$ &  $2.5^{+13.7}_{-2.1}$  \\
\enddata 
\tablecomments{Abundances are relative to \citet{anders89} Solar abundance.
	All errors are 90\% confidence for a number of interesting
	parameters equal to the number of free parameters in the model.
	The relative argon abundance is assumed to be equal to the neon
	relative abundance. The relative abundances of sodium, aluminum,
	calcium and nickel are assumed to be equal to the relative
	iron abundance. Model names are the same as used 
	in Table~\ref{tab:fits:multi_solar}.
	}
\tablenotetext{a}{At 90\% confidence this parameter is unconstrained within
	the range 0 to 50 $Z_{\sun}$.}
\end{deluxetable} 

The different CIE hot plasma models ({\sc Mekal}, 
{\sc Apec}, {\sc Equil} and the Raymond Smith code)
gave results consistent with each other in most
cases (the admittedly older-vintage Raymond Smith code gave the statistically
worst fits). If there are significant problems
with the atomic physics used in these codes then this affects all the
codes equally.

Using non-ionization equilibrium models does not lead to significantly
different metal abundances than the CIE models, so it does not appear that
the apparently low abundances are simply a consequence of unusual
ionization states in a single phase gas. 

\subsubsection{Multi-phase spectral model fits to the disk and halo}
\label{sec:spectra:multiphase}

\begin{figure*}[!ht]
\epsscale{2.0}
\plotone{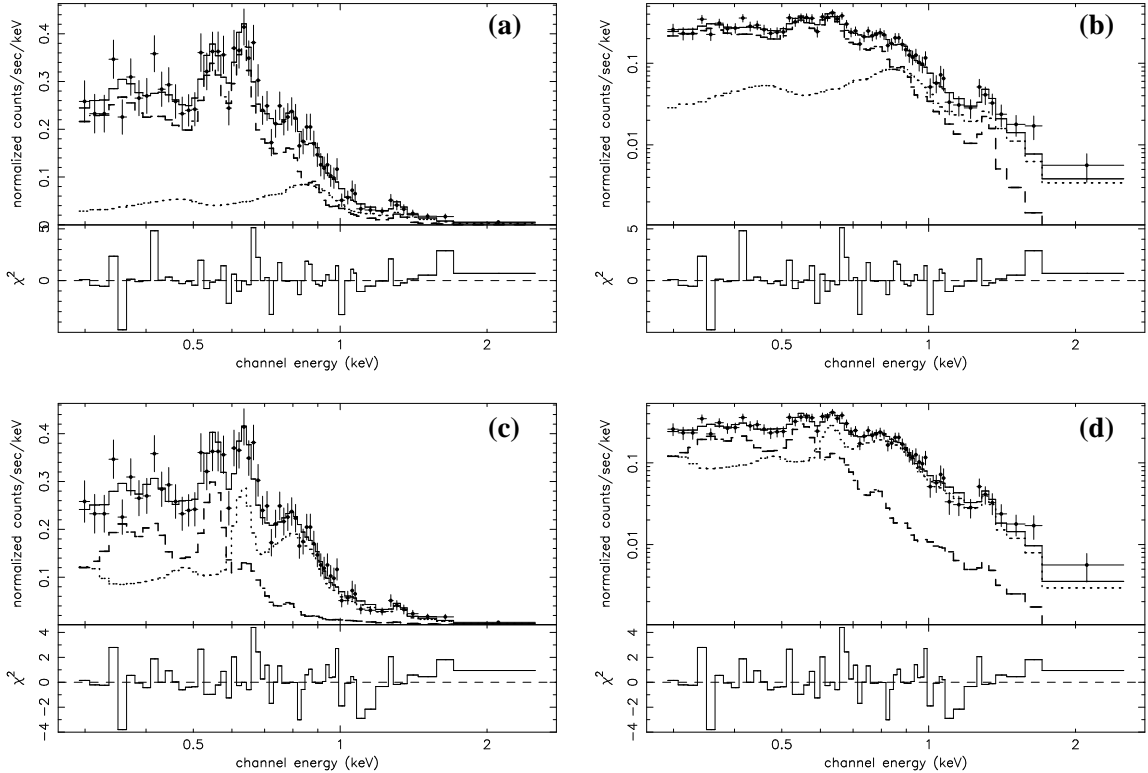}
\caption{The two best multi-phase spectral model fits to the
	halo spectrum. For details see Tables~\ref{tab:fits:multi_solar}
        \& \ref{tab:fits:metals:multi}. The
        {\tt 2Tvapec} model is shown on both linear and log intensity
        scales in panels (a) \& (b). The separate model components
        are shown as dashed lines.
        The {\tt 2Ivnei} model is show in linear and log form in
        panels (c) \& (d). The lower panel in each plot shows the
        contribution to the total $\chi^{2}$ from each data point.}
\label{fig:fit:multi_temp}
\end{figure*}

The strongest features in the halo spectrum are emission
at $E\sim 0.55$ and 0.63 keV (Fig.~\ref{fig:fit:multi_temp}), 
which are of roughly equal strength. 
Correcting for the $\sim 20$ eV error in the energy scale, 
these match up well with the O{\sc vii} and O{\sc viii} line complexes.
If both features come from the same gas phase, and collisional ionization
equilibrium holds, then the approximately equal strength of these
two features constrains the temperature to be $kT \sim 0.25$ keV.
The ratio of the O{\sc vii} line complex to O{\sc viii} line
complex count rates is a strong function of temperature (at $kT=0.15$ keV the
ratio is $\sim 5$, at 0.2 keV the ratio is $\sim 2$, at 0.25 keV the ratio is
$\sim 1$, at 0.3 keV the ratio is $\sim 0.5$ and at 0.4 keV the ratio is 
$\sim 0.25$).

This temperature is lower than the best-fitting temperature in the
single phase spectral models. Taken at face value, this forces
us to consider two phase or multi-phase spectral models in order to
obtain equal strength O{\sc vii} and O{\sc viii} emission in a spectrum
that is harder than a $kT = 0.25$ keV plasma. This argument should
be robust against calibration uncertainties, as it is
not based on the absolute intensities or detailed shape or
equivalent widths of the two features. Uncertainties in the 
effective area of the instrument are also unlikely to be significant
given the close proximity of the two features in energy space.

A multi-phase model can potentially explain the observed spectra
without requiring such low abundances. Multi-phase or multi-component
spectral models successfully removed the need to invoke extremely
low abundances in the {\it ROSAT} and {\it ASCA}
spectra of entire starburst galaxies (\citet{dwh98}; \citet{whd};
\citet{dahlem2000}). 
One phase with strong O{\sc vii} but weak O{\sc viii} emission,
along with another phase of higher temperature
having weak O{\sc vii} and strong O{\sc viii}, might look like a
$kT \sim 0.25$ keV plasma with lower oxygen abundance as the two
continuum components would reduce the equivalent width of each oxygen
line complex.

We fit a variety of two temperature CIE and NIE variable metallicity 
spectral models to the halo spectrum, 
using the {\sc Mekal}, {\sc Apec} and {\sc Nei} plasma codes.
For the multi-phase NIE models we considered cases where two components
of different temperature had the same ionization timescale (2Tvnei), or 
two components of different ionization timescale had the same 
temperature (the 2Ivnei model). 
We also considered two slightly more complex phase 
distributions, motivated by the hydrodynamical models of \citet{ss2000}
which predict that the X-ray emission integral as a function of
temperature is approximately a power law. In the PLT model the X-ray emission
integral is proportional to a power law function of the gas temperature
between a minimum and maximum temperature, \ie $EI(T) \propto T^{\alpha}$.
We also considered a Gaussian distribution in emission integral
as a function of (log) temperature, 
$EI(T) \propto \exp ( -4 \ln 2 \times [\{\log T - 
\log T_{\rm cent}\}/FWHM]^{2})$, where the Gaussian is centered
at temperature $T_{\rm cent}$.
In all cases we assumed that the multiple phases in the model had the
same metal abundance. The statistical quality of the data, along
with the current calibration uncertainties, do not justify 
multi-component fits where each component has a
different metallicity.
The results of these fits are presented in Tables~\ref{tab:fits:multi_solar}
\& \ref{tab:fits:metals:multi}.

Unfortunately, and somewhat to our surprise,
{\em multi-phase models do not solve the low metal abundance problem} in
the halo. Although the fits are statistically superior to those
using single temperature models, they remain somewhat poor
statistically-speaking. 
The best-fitting models (shown in Fig.~\ref{fig:fit:multi_temp}), 
the two temperature {\sc Apec}
and the two ionization timescale {\sc Nei} models have reduced
chi-squared values of 1.11 and 1.13 respectively. Such poor
values of reduced chi-squared would only occur $\sim 20$\% of the 
time if the models are accurately physical representations
of the emitting plasma.

The best-fit metal abundances are always extremely low, irrespective of
which model was used. In particular the best-constrained
elemental abundances, those of oxygen and iron, are consistently 
in the range 0.03 to 0.09 and 0.03 to 0.10 times Solar abundance
respectively. Admittedly the 90\% confidence regions for the oxygen abundance 
are moderately large, but given the systematically, unbelievably,
low best-fit abundances we have little confidence in the statistical
confidence regions.

Can depletion onto dust explain the apparently low gas-phase
abundances? Physically it is quite possible that the X-ray emitting gas may
contain a significant dust fraction. There is a wide range of 
evidence pointing to dusty material in superwinds
(\citet{phillips93}; \citet{alton99};
\citet{heckman2000}). It is striking that
FIR images of NGC 253 made by {\it ISOPHOT} at 60, 100 and $180\micron$
show enhanced dust emission along the X-ray-emitting arcs seen
in Fig.~\ref{fig:pspc_chips} \citep{radovich_isophot}.
Invoking depletion onto dust {\em does not} explain the fitted abundances,
where abundances of both refractory and non-refectory elements
are extremely low.
Although iron is heavily depleted onto dust in the warm ISM (by
factors of up to a hundred), oxygen is only weakly depleted
\citep{savagesembach96}.

Extremely low gas phase oxygen abundances are difficult to explain
physically, unless the X-ray emitting gas is relatively unenriched
material in the halo of NGC 253. For example some Galactic high 
velocity clouds have relatively low 
oxygen abundances, $Z_{O} \sim 0.1 Z_{O,\sun}$,
while other halo clouds possibly of Galactic origin have
approximately Solar oxygen abundance \citep{richter2001}.

\subsubsection{Spectral fits to the disk diffuse emission}

The hot
gas in the disk of NGC 253 can not be primordial, and therefore should
have oxygen abundances within a factor few of Solar.
If multi-phase spectral fits to the diffuse emission from the disk
also give strongly sub-Solar oxygen abundances then we will
have good evidence that the X-ray determined abundances can not be
trusted.

The disk diffuse spectrum was obtained from the 13ks Chandra observation
with the disk placed on the S3 chip. Based on the hardness ratio study
we excluded diffuse emission from 
starburst region and nuclear outflow cones, in addition to emission
from point sources. We included data from the spectrally soft region
to the east of the nucleus (see \S~\ref{sec:disk_hardness}). 
As the total flux from this region is
only a small fraction of the diffuse flux from the disk this will
not significantly complicate the spectrum of the disk.

We applied the same multi-phase spectral models to the disk diffuse
emission as had been used on the halo diffuse emission, the results of
which are shown for reasons of completeness 
in Tables~\ref{tab:fits:multi_solar}
\& \ref{tab:fits:metals:multi}. The most important result worth noting
is that both best-fit oxygen and iron abundances are again extremely
low in all the multi-phase models. 

This makes untenable the argument that low X-ray-derived oxygen abundances
arise from relatively primordial material suggested above (invoked
independently by \citet{xia2001} in the case of the extended
X-ray emission from the Ultra-luminous IR galaxy Mrk 273).

\subsubsection{A mix of thermal and non-thermal X-ray emission?}

The only method we have found to obtain reasonably high oxygen
and iron abundances is by adding a non-thermal component
to the spectral model (the PL+vapec model in 
Tables~\ref{tab:fits:multi_solar} \& \ref{tab:fits:metals:multi}). 
The thermal component provides the
observed line emission, while the non-thermal component
accounts for much of the continuum.

Non-thermal radio emission is seen within the haloes of
starburst galaxies such as M82 and NGC 4631 (\citet{ekers_n4631radio};
\citet{seaquist91}), including NGC 253 \citep{carilli_n253radio},
and is thought to be synchrotron emission from cosmic ray (CR) electrons
that arise within within the disk (and are advected out within the
superwind), or are accelerated locally in internal wind shocks.


The modeled non-thermal component represents a relatively fixed
fraction of the total emission at most energies, and only clearly 
dominates the emission at energies $E > 1.5$ keV. The soft
X-ray images presented in Figs.~\ref{fig:halo_diffuse} 
\& \ref{fig:pspc_chips} therefore represent a reasonable
approximation to the spatial distribution of the hypothesized
non-thermal component, which we might expect to look similar the
the non-thermal radio emission.
An overlay of the soft X-ray {\it ROSAT} PSPC image over the radio
emission is presented in \citet{pietsch2000}. The north-eastern
and south-eastern arcs of X-ray emission seen by the PSPC
do match up to relatively distinct features in the 330 Mhz radio
emission --- the ``prominence'' and the ``spur'' respectively
of \citet{carilli_n253radio}. Nevertheless, in general the radio emission
is dissimilar to the X-ray emission, extending to larger galactocentric
radio along the plane of the galaxy and within the lower halo
immediately above the disk, at the same radio surface brightness as within
the X-ray-bright regions of the halo.

Given general lack of similarity between X-ray and radio emission,
along with the {\it Chandra} calibration uncertainties 
and the relatively poorly
constrained  parameters of the fitted spectral model, we do not
believe mixed thermal plus non-thermal X-ray emission models
currently provide a likely solution to the metal abundance problem. 
Mixed thermal and non-thermal X-ray emission is a
possibility worth bearing in mind for the future, and requires more
theoretical work to explore the physical feasibility of
generating relatively bright synchrotron or IC X-ray emission.

\subsubsection{A summary of the abundance puzzle}

We note low X-ray-derived metal abundances appear to be a common feature
in {\it Chandra} ACIS spectra of the 
extended diffuse emission in starbursting galaxies
(\eg NGC 4631 \citep{wang2001}, Mrk 273 \citep{xia2001}),
in addition to the longer standing findings from {\it ROSAT} \& {\it ASCA}.
ACIS spectra of objects that are not starbursts do not give such
low best-fit element abundances\footnote{\citet{kastner2001} 
report near-solar O \& Ne abundances, and super-Solar He, C, N, Mg \& Si in a
{\it Chandra} ACIS-S study of the diffuse X-ray emission from
the planetary nebula NGC 7027. \citet{hughes2000} find super-Solar
S and Si abundances in the SNR Cassiopeia A.}. 

Neither
dust depletion, nor naturally low abundance material, nor 
fitting problems due to thermal emission from
multi-phase gas appear to solve this problem.
Calibration problems are a likely cause of the moderately high
best-fit $\chi^{2}$ values, but we do not believe that they
are responsible for the factor $\sim 20$ lower-than-Solar oxygen 
abundances found in spectra of starbursts. 

A mixed thermal plus non-thermal emission model gives more normal abundances
in the halo and the disk gas.
Although mixed thermal and non-thermal emission is a physically possible 
solution to the metal abundance problem, it requires
something of a spectral conspiracy to so effectively hide a non-thermal
spectrum under some thermal emission in the both the halo
and the disk.

The observed ACIS spectra of both the
halo and the disk look empirically like emission from low metal abundance
hot plasmas. Given the line
features in the spectra there is no doubt that there is some
form of thermal plasma in both the halo and the disk.
The limited number of spectral diagnostics
that can be applied to the spectra appear consistent with each other.
Physically the low abundances are 
unreasonable, and we are forced to conclude that
we do not understand in detail the X-ray spectra of diffuse emission
from star-forming galaxies. The problem of X-ray-derived metal
abundances in starbursts uncovered by {\it ROSAT} and {\it ASCA}
is not yet solved, and will require an imaging spectroscope with
the spectroscopic resolution sufficient to robustly measure true
line equivalent widths {\em and} the spectral shape of the continuum. 
We continue to urge readers to treat with caution any
claims made on the basis of X-ray derived metal abundances in
star-forming galaxies.

\subsection{Physical properties of the X-ray-emitting plasma}
\label{sec:spectra:properties}

\subsubsection{The soft thermal emission}

All spectral models give very similar values for the
X-ray flux in the 0.3 -- 2.0 keV energy band,
$f_{\rm X} \approx 5.0 \times 10^{-13} \ergps \pcmsq$.
This value is relatively robust, as it is closely
related to the directly observed count rate.
Absorption-corrected fluxes are more model-dependent,
and also more uncertain due to both statistical
and systematic calibration uncertainties that affect
the best-fit hydrogen column (see \S~\ref{sec:data:calibration}).
The 2Tvapec model absorption corrected 0.3 -- 2.0 keV
X-ray flux for the northern halo is $f_{\rm X} = 7.9 \times 10^{-13}
\ergps \pcmsq$ ($^{+210\%}_{-64\%}$), while the absorption-corrected
flux in the 2Ivnei model is $f_{\rm X} = 6.0 \times 10^{-13} \ergps
\pcmsq$ ($^{+258\%}_{-64\%}$).
Errors are
based on 90\% confidence
regions for each model's normalization.
Adopting the latter estimate yields a total X-ray luminosity of
$4.9\times 10^{38} \ergps$ for the superwind in the
northern halo region

Plasma properties such as densities, masses and energy content
are much more uncertain. {\em The following estimates are useful
as order-of-magnitude estimates only.} We have used the total
emission integral from the 2Tvapec model, $EI = 3.7 \times 10^{62} \pcc$,
and a mean temperature of $T \sim 3 \times 10^{6} \K$
in the following estimates. Note that the 2Ivnei model and
the thermal component in the PL+vapec model give
total thermal plasma emission integrals one order of magnitude lower,
which would reduce all plasma properties below (except the mass cooling
rate) by a factor $\sim 3$.

The unknown volume filling factor
$\eta_{\rm X}$ of the X-ray emitting gas enters into calculations
of the electron number density based on the model-derived emission
integral $EI = n_{\rm e} n_{\rm H} \eta_{\rm X} V$.
Possible values of $\eta_{\rm X}$ lie in the range $10^{-3}$ to 1.

As we do not believe the best-fit metal
abundances, we explicitly include the scaling factor
${\cal R} = 0.03/Z_{\rm O, true}$ 
in the following estimates as a correction to the
derived emission integrals\footnote{While the filling factor
dependence of superwind plasma properties is often 
explicitly noted in published work, the metallicity dependence
is generally not shown. The plasma properties of the various 
superwinds given in the sample
\citet{rps97} assumed a variety of very low abundances.
\citet{sps97} assumed $Z = 0.05 Z_{\sun}$ for M82, while \citet{pietsch2000}
assumed Solar abundance in NGC 253's halo.}.
We believe a reasonable value for the true oxygen 
abundance $Z_{\rm O,true}$ is $\sim 0.5$, if the X-ray-emitting plasma
is ambient disk gas dragged into the halo by the superwind.

Treating the halo region as cylinder of diameter 6.35 kpc and height
6.35 kpc gives a total volume of $V = 5.91 \times 10^{66}$ cm$^{3}$.
The electron density in the X-ray emitting plasma in the halo is 
$n_{\rm e} = 7.9 \times 10^{-3} \eta^{-0.5}_{\rm X} {\cal R}^{0.5} \pcc$.
The thermal pressure $P_{\rm TH}/k = 4.7\times10^{4} \eta_{\rm X}^{-0.5}
{\cal R}^{0.5} \K \pcc$.
The mass of the this gas is $M_{\rm X} = 4.7\times10^{7} \eta_{\rm X}^{0.5}
{\cal R}^{0.5} \Msol$, its cooling time and mass cooling rate
$t_{\rm cool} = 770 \eta_{\rm X}^{0.5} {\cal R}^{0.5}$ Myr 
and $\dot M_{\rm X} = 0.06 \Msol \pyr$ ($\dot M_{\rm X}$ depends
only on the gas temperature and bolometric luminosity).
The mass flow rate $\dot M_{\rm flow} 
\sim M_{\rm X} v_{\rm x}/z = \rho A v_{\rm x} \sim 7.6 \eta_{\rm X}^{0.5}
{\cal R}^{0.5} v_{\rm 1000} \Msol \pyr$, where $v_{\rm 1000}$
is the outflow velocity of the X-ray emitting plasma in units of
1000 km/s, and we have taken the vertical height of the region $z$
to be 6.35 kpc. The thermal and kinetic energy content of this
plasma are $E_{\rm TH} = 5.9 \times 10^{55} \eta_{\rm X}^{0.5} 
{\cal R}^{0.5} \erg$ and $E_{\rm KE} = 4.7 \times 10^{56} 
\eta_{\rm X}^{0.5} {\cal R}^{0.5} v_{\rm 1000}^{2} \erg$.

\subsubsection{Upper limits on a high temperature diffuse component}

As discussed previously,
the detection of counts in the hard X-ray
energy band, of marginal significance,
 is most probably due to systematic problems
with background subtraction. If we treat the 2 -- 8 keV
count rate as a conservative upper limit on the emission from a $kT = 4$ keV
plasma component in the superwind, the 2 -- 8 keV
X-ray flux, luminosity and total emission integral from very hot gas in this
region of the superwind are 
$f_{\rm X,hot} < 1.4 \times 10^{-13} \ergps \pcmsq$,
$L_{\rm X,hot} < 1.1 \times 10^{38} \ergps$,
and $EI_{\rm hot} < 1.4 \times 10^{60} \pcc$.

This is consistent with
numerical models of superwinds (\citet{suchkov94}; \citet{ss2000}), which
predict that although plasma at such temperatures exists within superwinds,
it is of such low density that it does not radiate effectively.

\section{Point sources and compact clouds in the halo}
\label{sec:halo_point_sources}

We detect 30 point-like sources in the S3-chip pointing on the northern
halo, that have $S/N \ge 2$ in any of the three broad, soft or hard
energy bands (see Table~\ref{tab:point_sources}). 
Only three of these sources were detected in previous
{\it ROSAT } PSPC and HRI observations, which were at least 1.5 orders
of magnitude less sensitive than our 43 ks {\it Chandra} observation.

Fig.~\ref{fig:allchips} shows that these sources appear to cluster
along the arc of diffuse emission. Could some of them
be  X-ray-bright clumps or clouds in the
wind, as has been suggested on several occasions for some 
{\it ROSAT}-detected sources in superwinds (see \citet{read94}; 
\citet{sps97}; \citet{vp99})?

\subsection{The nature of the point-like sources}

We have found optical counterparts for 12 of these sources 
(Table~\ref{tab:optical_srcs}). The high X-ray-to-optical flux ratio
of most of these sources are consistent with them being AGN 
(see \citet{krautter99}),
although two of the sources (numbers 9 \& 16) are without
doubt stars given their low X-ray-to-optical flux ratios..

The total number of sources detected in the soft (19) and hard bands (11)
with $S/N \ge 2$ is consistent with the numbers expected 
from deep {\it Chandra} surveys. The $N \log S$ prescription
given in \citet{giaconni2001} predicts $18\pm5$ and $10\pm4$
sources in the soft and bands over the entire S3 chip,
for the flux limit reached in our observation. This suggests that 
the sources we detect that do not have optical counterparts 
are nevertheless genuine point-like X-ray sources, primarily
AGN, and not false detections of compact
clumps in the diffuse X-ray emission.

The spectral properties of these sources (with the exception 
of the two known stars) are consistent with them being AGN or
obscured AGN. Most of these sources are individually too faint to
obtain robust hardness ratios from, let alone spectra, but the
emission from many sources can be combined to obtain a composite 
spectrum. As we are interested in whether the faintest sources
may be compact features in the diffuse emission we
extracted a composite ACIS spectrum from all sources with $2 \le S/N \le 3$
that did {\em not} have any optical counterpart. We refer to these
sources as the faint sources, and the composite spectrum as the 
faint source spectrum. A composite spectrum of the brighter
sources ($S/N \ge 3$), and in addition 
those sources with optical counterparts,
but excluding the two known stars, was also extracted (the bright
source spectrum). Finally, we also extracted a composite spectrum of the
two stars, which we expect to be spectrally softer than the other
sources. Source number 20 (a
known $z=1.25$ AGN, see \citet{vp99}), is sufficiently bright enough to
obtain a spectrum from.

An absorbed power law spectral model fit to the bright source composite
spectrum gives a best fit photon index of 
$\Gamma = 1.44^{+0.11}_{-0.12}$ (90\% confidence
in three interesting parameters), absorbed by a column 
of $\nH = 1.5^{+1.9}_{-1.3} \times 10^{20} \pcmsq$ 
($\chi^{2}_{\nu} = 1.34$). This is
a harder spectrum than a typical AGN ($\Gamma \sim 1.7$), but
is consistent with the mean spectrum of the hard X-ray background.
The X-ray spectrum of source number 20 is consistent with a typical
AGN, the best-fit model giving $\Gamma = 1.66^{+0.30}_{-0.23}$
and $\nH = 4.9^{+3.9}_{-2.4} \times 10^{20} \pcmsq$ 
($\chi^{2}_{\nu} = 1.08$).

Hardness ratios of the bright and faint source spectra are shown
in Table~\ref{tab:point_hardness_ratios}.
The energy bands used  (soft: 0.3 -- 1.0 keV,
hard: 1.0 -- 8.0 keV) was chosen
to give the largest difference in hardness between the bright sources
and the large-scale diffuse emission in the halo. If a significant fraction
of the faint sources are indeed dense clumps in the wind, we would
expect the hardness of the composite faint source spectrum to be softer than
the bright sources, and similar to the hardness ratio of the
diffuse emission. 
This is not the case -- the hardness of the faint
sources is statistically identical to that of the bright sources,
and differs from the hardness of the diffuse emission by at least 
$6\sigma$. The two X-ray sources identified as stars based on
X-ray-to-optical flux ratios are spectrally soft using this hardness 
ratio test, confirming the identification of the X-ray sources
with the optical stars.

Thus it appears that the faint sources seen in the halo
of NGC 253 are indeed
background AGN and foreground stars
unrelated to the superwind itself, based
both on number counts and on their spectral properties.

\begin{deluxetable}{lrr}
 \tabletypesize{\scriptsize}%
\tablecolumns{3} 
\tablewidth{0pc} 
\tablecaption{Point source hardness ratios
	\label{tab:point_hardness_ratios}} 
\tablehead{ 

\colhead{Component} 
	& \colhead{$Q_{1-8,0.3-1}$\tablenotemark{a}} 
	& \colhead{Count rate\tablenotemark{b}}
	}
\startdata
Bright sources       & $0.109\pm{0.028}$  & $0.0379\pm{0.0010}$ \\
Source 20\tablenotemark{c} & $0.077\pm{0.064}$  & $0.0113\pm{0.0006}$ \\
Faint sources        & $0.138\pm{0.128}$  & $0.0050\pm{0.0006}$ \\
Stars                & $-0.580\pm{0.194}$ & $0.0037\pm{0.0006}$ \\
Diffuse halo emission  & $-0.678\pm{0.031}$ & $0.2092\pm{0.0053}$ \\
\enddata 
\tablenotetext{a}{Hardness ratio using the number of
  counts in the 1.0 -- 8.0 keV energy as the hard band and
  the counts in 0.3 -- 1.0 keV energy
  band as the soft band. Errors are 68\% confidence.}
\tablenotetext{b}{Total ACIS-S3 background-subtracted count rate in 
	the 0.3 -- 8.0 keV energy band.}
\tablenotetext{c}{Source number 20 is included in the composite bright source
	spectrum.}
\end{deluxetable}


\begin{figure*}[!ht]
\epsscale{2.0}
\plotone{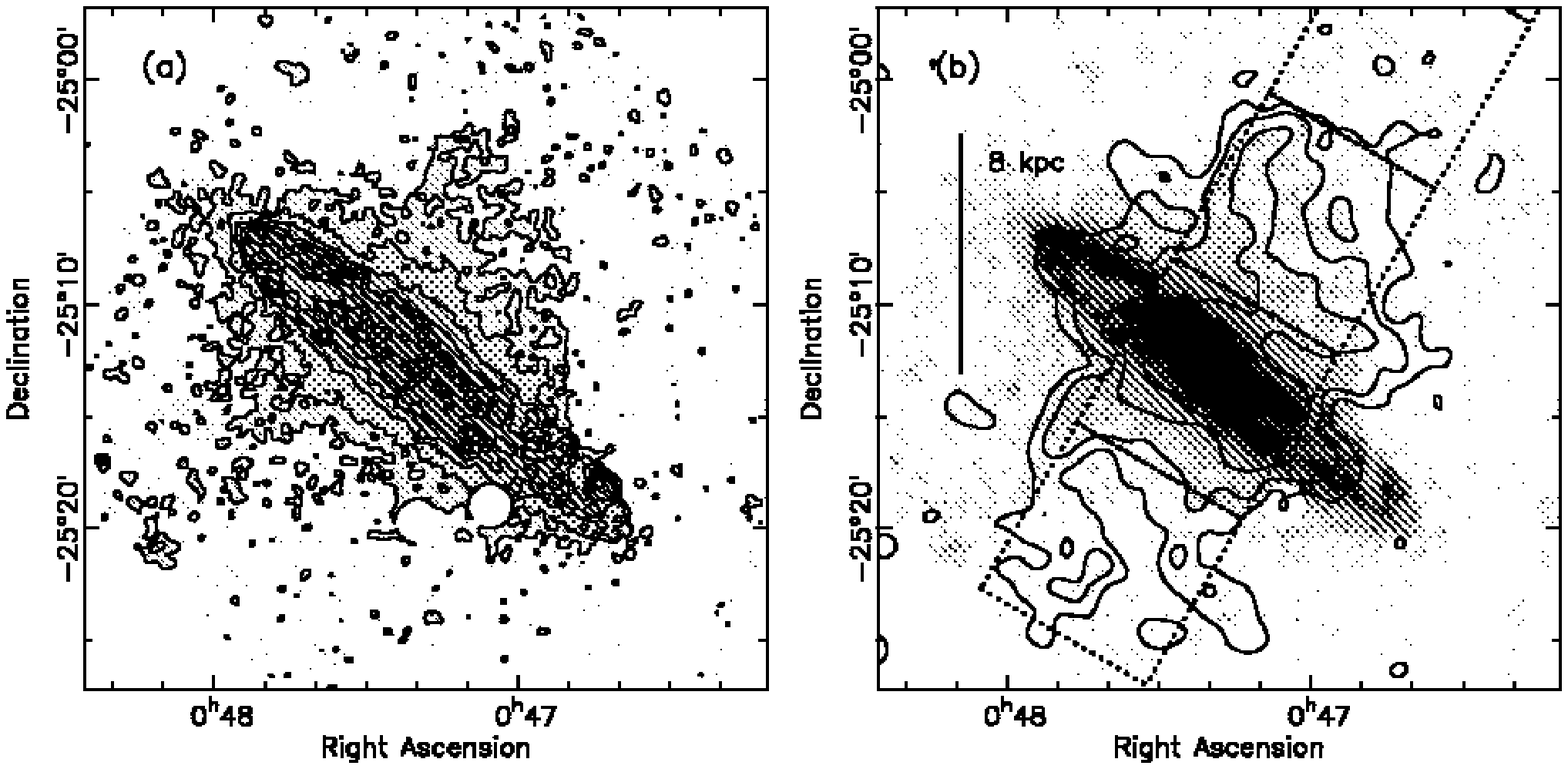}
  \caption{(a) Continuum subtracted \halpha~image of NGC 253 
	showing emission associated with the
	superwind in the halo of the galaxy. The original 
	image (published in \citet{hoopes96}) has been smoothed
	with a Gaussian mask of $FWHM = 20\arcsec$ in order to bring
	out the very low surface emission in the halo, and is shown
	using a logarithmic intensity scale between $10^{-18}$ and
	$3.9 \times 10^{-15} \ergps \pcmsq$ arcsec$^{-2}$. Contours
	start at a \halpha~surface brightness of $2\times10^{-18}
	\ergps \pcmsq$ arcsec$^{-2}$ and increase in steps of 0.5 dex.
	Residuals from two saturated star images 
	to the south of the galactic disk have been excised
	from the image.
	(b) The \halpha~image overlaid with the contours of soft X-ray
        emission seen by the {\it ROSAT} PSPC (Fig.~\ref{fig:pspc_chips}b).
        Dotted lines show the location of the ACIS chips in our 
   	{\it Chandra} observation. The \halpha~image is shown on
	a logarithmic scale between $10^{-18}$ and 
	$2\times 10^{-16} \ergps \pcmsq$ arcsec$^{-2}$.
  }
  \label{fig:halpha_images}
\end{figure*}

\begin{figure*}[!ht]
\epsscale{2.0}
\plotone{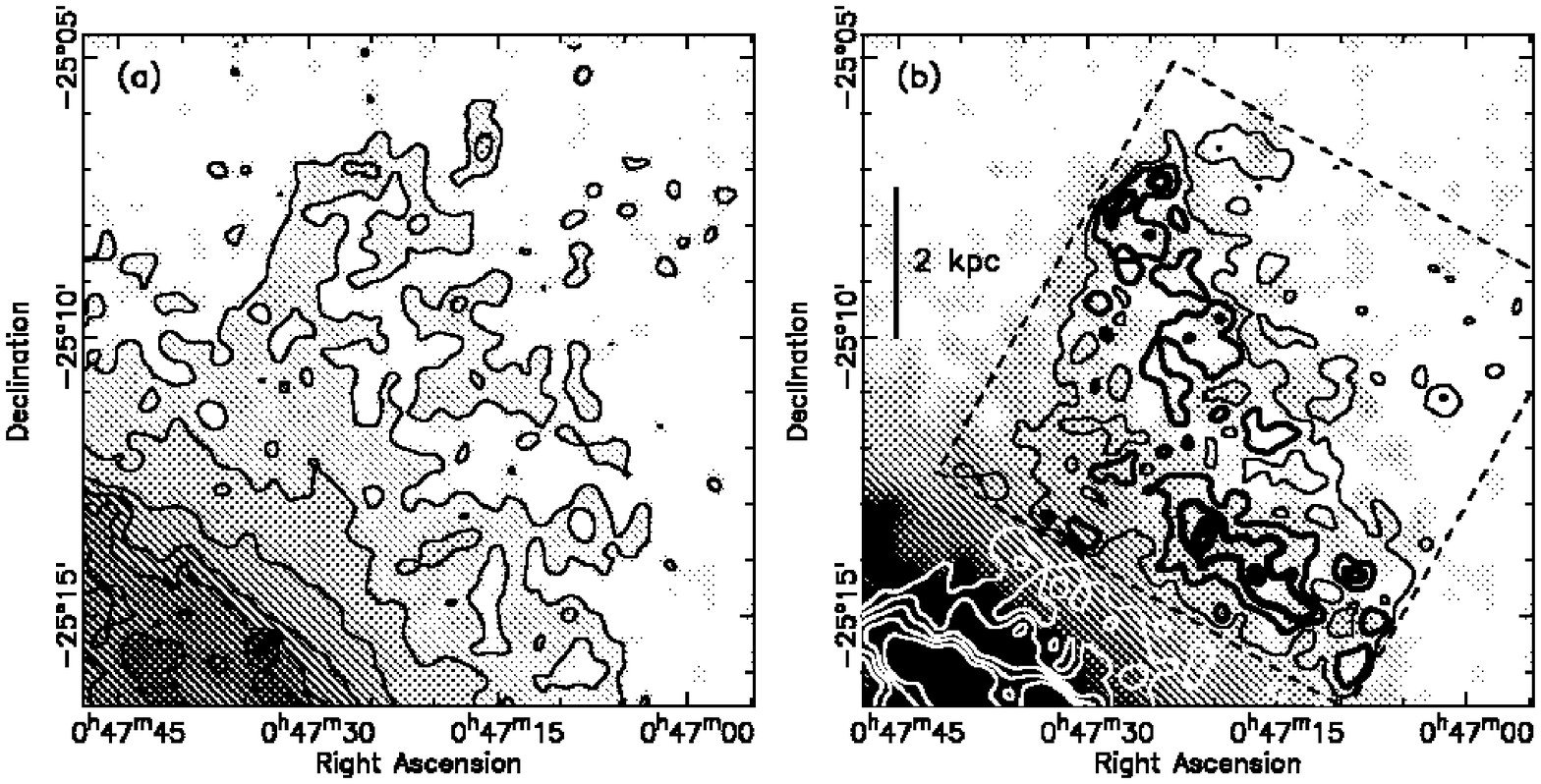}
  \caption{(a) As Fig.~\ref{fig:halpha_images}a, except focusing in
	on the northern halo region to better show the arcs and filaments of
	\halpha~emission.
	(b) As  Fig.~\ref{fig:halpha_images}b, but focusing in on the
	northern halo region to better show the relationship between the
	\halpha~and soft X-ray emission using the $20\arcsec$-smoothed
	{\it Chandra} ACIS data, 
 	instead of the lower resolution and sensitivity PSPC data.
	The contours are X-ray S/N = 2, 3, 4 \& 5 
	(see Figs.~\ref{fig:halo_diffuse}b \& \ref{fig:disk_hardness}b),
	shown with progressively bolder solid lines. The dashed line
	represents the edges of the S3 chip.
	}
  \label{fig:halpha_zoom}
\end{figure*}

\section{\halpha~and X-ray emission in the halo}
\label{sec:results:halpha}

The relationship between soft X-ray emission and optical \halpha~emission,
if present, provides strong clues to the origin of the X-ray-emitting
gas in superwinds. It has been long known that there is a strong
correlation between X-ray and \halpha~emission in the bright
central $\sim 1$ kpc regions of superwinds (\eg \citet{watson84};
\citet{mccarthy87}), robustly confirmed by 
{\it Chandra} observations in the case of NGC 253 \citep{strickland2000}. 

For several starburst systems with outflows the thermal
X-ray and \halpha~emission correlate well on larger $\sim 10$ kpc scales
(\eg Arp 220 \citep{heckman_arp220}, M82 \citep{lehnert99}), 
but no 10 kpc-scale \halpha~emission 
has been previously noted in the halo of NGC 253, despite its proximity
and status a well-studied archetypal starburst.

Nevertheless, faint \halpha~emission, very similar in morphology to the
diffuse X-ray emission, does exist in the halo of NGC 253 
(Figs.~\ref{fig:halpha_images} \& \ref{fig:halpha_zoom}).
To the north of the disk this correlation between X-ray and 
\halpha~emission is particularly clear. 
In the northern halo the X-ray emission
appears to peak in surface brightness slightly to the west
of the peak in the \halpha~emission, although it is difficult to
be absolutely confident of this due the small
{\it Chandra} ACIS field of view (Fig.~\ref{fig:halpha_zoom}), 
and the heavy smoothing required by the {\it ROSAT} PSPC data.
(Fig.~\ref{fig:halpha_images}).
On the southern side of the galaxy, south east of the
nucleus, a shorter X-ray arc coincides almost exactly with an
\halpha~arc with no discernible spatial offset.
Although the morphology	
	of the \halpha~and X-ray emission is very similar on $\sim 5$ kpc
	scales, 
	in detail there does not appear to be a exact
	correlation or anti-correlation on $\sim 1$ kpc scales.

\subsection{H$\alpha$/X-ray flux ratios}
\label{sec:results:halpha:halo}

The surface brightness of \halpha~emission in the halo of
 NGC 253 is $\sim2$ orders of magnitude lower
than the typical \halpha~surface
brightness in the disk.
Our best estimate of the mean surface brightness of the northern halo
\halpha~emission is $7.72 \times 10^{-15} \erg \ps \pcmsq$ 
arcmin$^{-2}$, in the
70.5 arcmin$^{2}$ halo region covered by the ACIS S3 chip. 
The \halpha~luminosity in the halo field
is $L_{H\alpha} \sim 4.4 \times 10^{38}$ \ergps,
comparable to the  total luminosity of the soft
X-ray emission.

The \halpha~to X-ray flux ratio in this diffuse emission
is interesting, as it may provide clues to the physical
mechanisms behind both the origin of the X-ray emitting gas
and the ionization source of the cool \halpha~emitting gas.
Previous work has shown that the soft X-ray and \halpha~fluxes
are roughly comparable\footnote{\citet{shopbell98} quote 
$f_{\rm H\alpha}/f_{\rm X} \sim 30$ for the brightest \halpha~filaments 
in M82 based on the {\it ROSAT} HRI observations. 
This may be a beam-smearing artefact, given 
to the much poorer spatial resolution of the
HRI compared to their \halpha~observations.} 
in both the M82 \citep{lehnert99} and Arp 220 \citep{heckman_arp220}
superwinds.

\halpha/X-ray flux ratios for the northern halo region and for 
the clearly-limb-brightened region of the nuclear southern
outflow cone \citep{strickland2000} are given in 
Table~\ref{tab:halpha_xray_fluxratios}. Flux ratios
are also given for the four northern halo quadrants we defined,
(see Fig~\ref{fig:allchips}b) which probe a mix of
X-ray and \halpha~bright and faint regions. 

It is interesting that in all these regions the \halpha/X-ray
flux ratio is generally within a factor 2 of unity.
Even in the
north western region of the halo, where there is little X-ray and
little \halpha~emission, the fluxes are closely matched 
($f_{\rm H\alpha}/f_{\rm X} = 0.44$).

\begin{deluxetable}{lrrrrrr}
 \tabletypesize{\scriptsize}%
\tablecolumns{7} 
\tablewidth{0pc} 
\tablecaption{H$\alpha$/X-ray flux ratios and surface brightnesses
	\label{tab:halpha_xray_fluxratios}} 
\tablehead{ 

\colhead{Region} & \colhead{Area} 
	& \colhead{$f_{\rm H\alpha}$}
	& \colhead{$f_{\rm X}$} 
	& \colhead{$\Sigma_{\rm H\alpha}$}
	& \colhead{$\Sigma_{\rm X}$}
	& \colhead{$f_{\rm H\alpha}/f_{\rm X}$} \\
\colhead{(1)} & \colhead{(2)} & \colhead{(3)} 
	& \colhead{(4)} & \colhead{(5)} 
	& \colhead{(6)} & \colhead{(7)}
	}
\startdata
Nuclear cone\tablenotemark{a} &   0.27 & 1.81 & 1.79 & 186.1 & 184.2 & 1.01 \\ 
Halo (total)                  &  70.51 & 5.44 & 5.03 &   2.1 &   2.0 & 1.08 \\ 
NW halo                       &  17.63 & 0.27 & 0.61 &   0.4 &   1.0 & 0.44 \\ 
NE halo                       &  17.63 & 1.00 & 1.26 &   1.6 &   2.0 & 0.79 \\
SW halo                       &  17.63 & 1.41 & 1.79 &   2.2 &   2.8 & 0.79 \\ 
SE halo                       &  17.63 & 2.76 & 1.37 &   4.4 &   2.2 & 2.01 \\ 
\enddata 
\tablecomments{Column 1: Spatial region. 
	(2): Geometrical area of region in arcmin$^{2}$.
	(3): Estimated \halpha~flux in $10^{-13} \ergps \pcmsq$.
	(4): Estimated soft X-ray flux (0.3--2.0 keV energy band) 
	in $10^{-13} \ergps \pcmsq$, corrected for
        the area lost due to point sources.
        This flux estimate has not been corrected for absorption,
        and is robust across all spectral models. Absorption-corrected
        fluxes are model dependent, \eg 
	20\% higher for the 2Ivnei model, and 60\% higher
        for the 2Tvapec model (Table~\ref{tab:fits:multi_solar}).
        Columns (5) \& (6): Mean H$\alpha$ (5) and soft X-ray (6) 
	surface brightnesses in units of $10^{-18}$ erg s$^{-1}$ 
	cm$^{-2}$ arcsec$^{-2}$.}
\tablenotetext{a}{The clearly limb-brightened region of the southern
	nuclear outflow cone discussed in \citet{strickland2000}.}
\end{deluxetable}

\section{Discussion}
\label{sec:discussion}

\subsection{Models for X-ray and H$\alpha$ emission in superwinds}
\label{sec:discussion:models}

The general correlation between the soft thermal X-ray emission and 
optical line emission in the halo of NGC 253 (Fig.~\ref{fig:halpha_images})
strongly suggests that these two forms of emission are physically
linked, and any meaningful model of superwinds must explain this
spatial relationship in addition to the relative fluxes between the two.

\begin{figure*}[!ht]
\epsscale{2.0}
\plotone{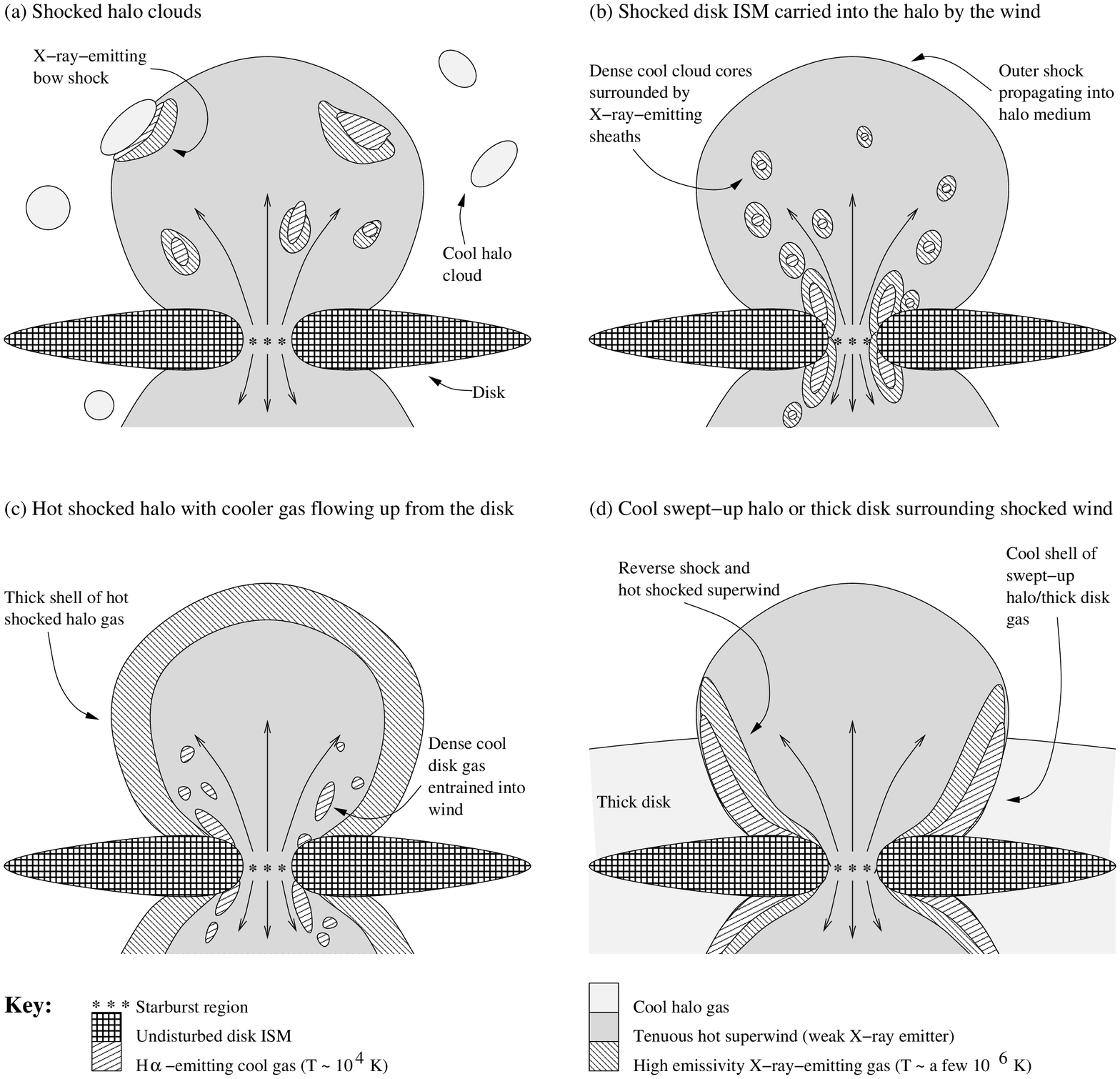}
\caption{A schematic representation of some of the possible
  situations giving rise to spatially-correlated X-ray and 
  \halpha~emission in the haloes of starburst galaxies,
  shown as imaginary 2-dimensional slices through the plane
  described by the major and minor axes of the galaxy.
  Only one pole of the bipolar outflow is shown in detail. 
  See \S~\ref{sec:discussion:models} for details.}
\label{fig:cartoon}
\end{figure*}

In this section we shall assess the physical plausibility of
 a variety of possible mechanisms for
producing both \halpha~and soft X-ray emission in the halo of NGC 253,
based on previously  published discussions of supernova remnants,
wind-blown bubbles, superbubbles and superwinds.
A cartoon depiction of some of these possible scenarios
is presented in Fig.~\ref{fig:cartoon}, in the hope of making following
discussion clearer to the reader.

\subsubsection{Sources of H$\alpha$~emitting gas}

There are a variety of possible sources of the cool gas responsible
for the \halpha~emission in the halo. An unlikely possibility is that
it is cooled superwind SN-ejecta. In-situ halo sources of \halpha-emitting
gas are a radiatively-cooled shell of swept-up halo gas, or radiative shocks
driven into pre-existing clouds (\eg \citet{hrb94}) within the halo.

The ionization source may be photoionization or shock-heating.
Photoionization may be either by UV photons
propagating up into the halo from the starburst, or by local sources
of energetic photons from the superwind.

\subsubsection{Sources of X-ray emitting plasma}

Possible sources of X-ray emitting plasma are shocked 
\citep{cc} or mass-loaded (\citet{suchkov96}; \citet{hartquist97})
superwind material, a volume-filling halo medium swept-up and 
shock-heated by the superwind, or interaction between the wind
and clouds in the halo (either pre-existing halo clouds or disk
gas dragged into the halo by the wind).

\subsubsection{Possible emission models for the halo of NGC 253}

\subsubsubsection{Model 1: Cooling radiation from the wind}

Could both the X-ray and \halpha~emission come from cooling of
the volume-filling wind fluid? The advantages of this
model are its simplicity (in that it does not depend on the 
presence dense gas in the halo), and that it offers the attractive
possibility that observations directly probe the enriched gas
that drives the wind.

This model requires that the cooling time in the wind is less than,
or of order, the age of the outflow. The dynamical age of the gas
at the center of the halo region is $t_{\rm dyn} \sim 6 v^{-1}_{1000}$ Myr,
where $v_{1000}$ is the mean outflow velocity in units of 1000 km/s.
The {\em minimum} reasonable values of $v_{1000}$ are between 0.3 -- 0.6
(see the more detailed discussion of outflow velocity in Model 2 below),
which gives dynamic ages of 10 to 20 Myr, 
similar to estimates of the age of the starburst
\citep{engelbracht1998}.
The {\em radiative} cooling time of the 
X-ray-emitting gas is $t_{\rm cool} \sim
770 \eta_{\rm X}^{0.5} {\cal R}^{0.5}$ Myr 
(\S~\ref{sec:spectra:properties}), significantly longer than
the dynamical age of the outflow for any reasonable X-ray filling
factor or metal abundance.

\citet{breitschwerdt99} propose a model for soft
X-ray emission in 1-dimensional cosmic-ray-driven galactic outflows where the
X-ray emitting plasma has a temperature of a few $\times 10^{4}$ K,
due to adiabatic cooling. The X-ray emission arises due to delayed
recombination, in which the ions have not had time to come into
ionization equilibrium and hence are over-ionized for their kinetic
temperature. This neatly bypasses the radiative cooling time problem
discussed above.

However, this model can not explain the \halpha~emission
seen in superwinds.
Although \citet{breitschwerdt99} do not consider 
\halpha~emission, it seems likely that recombination of electrons
onto protons at $T \sim$ a few $\times 10^{4}$ K 
would lead to spatially correlated \halpha~and
soft X-ray emission in this model. However, it is unlikely that the plasma
in starburst-driven superwinds expands such that adiabatic cooling
leads to low kinetic temperatures. Multi-dimensional
hydrodynamical simulations (\eg \citet{tomisaka88}; \citet{tomisaka_bregman};
\citet{suchkov94}; \citet{ss2000})
accurately follow the kinetic temperature
of the plasma and
all show that internal shocks are extremely common in superwinds, reheating
the plasma as kinetic energy is converted into thermal energy in the
shocks. This rules out strong adiabatic cooling of the
wind giving rise to the \halpha~emission, and probably rules out 
the specific delayed recombination model in 
\citet{breitschwerdt99}\footnote{We agree whole-heartedly 
with the more general point
\citet{breitschwerdt99} make, which is that collisional 
ionization equilibrium is unlikely to apply in 
many regions of galactic outflows.}
from applying to starburst-driven superwinds.

Another significant 
problem with this model, where the emission is distributed 
smoothly throughout the wind, is that it does not explain
X-shaped morphology of both the X-ray and \halpha~emission.

Hence the model of the X-ray and \halpha~emission as coming from the
cooling of the volume-filling wind-fluid can 
be confidently rejected.

\subsubsubsection{Model 2: Shocked clouds in the halo}

A more plausible model is one developed by \citet{lehnert99}
to explain an X-ray and \halpha~emitting ridge $\sim 12$ kpc
above M82: a pre-existing neutral cloud (most probably
tidal debris associated with the M82/M81 interaction) is over-run
by the expanding superwind. A slow shock is driven into the
cloud, ionizing it. Photoionization by UV photons escaping from the
starburst region may also contribute to the ionization of the cloud.
A stand-off bow shock of hot X-ray-emitting
superwind gas forms upstream of the cloud. The net result is
a close spatial correlation between \halpha~and X-ray emission,
both of which are only visible in the vicinity of any halo cloud
(see Fig.~\ref{fig:cartoon}a). 

We refer the reader to \citet{lehnert99} for a detailed
discussion of this model. We shall only elaborate on some details
of the feasibility of the model as applied to the halo of NGC 253
here.

Shock-heating of the wind material in the cloud bow shock to achieve 
a temperature of $T = 3 \times 10^{6}$ K requires a shock velocity of
$v_{\rm s}= [(16/3) \thinspace (kT/\mu)]^{0.5} \approx 470$ km/s.
Assuming radiative losses in the starburst region are insignificant,
the outflowing SN-ejecta could have velocities as high as $3000$ km/s.
Spectroscopic measurements of the \halpha~emitting gas in the
southern nuclear outflow cone give deprojected outflow velocities
of between 340 to 390 km/s (\citet{ham90}; \citet{schulz92}) based on
an opening angle for the cone of 50 to $65\degr$. Smaller
opening angles (\eg \citet{strickland2000}) imply the 
deprojected \halpha~outflow
velocity may be $\sim 600$ km/s. The SN-ejecta which sweeps
up and accelerates the \halpha~emitting gas is likely to have higher
outflow velocities, so the observed temperature of the X-ray
emitting gas in the halo NGC 253 is easily produced in shocks.

The spectral uniformity of the diffuse X-ray emission is a natural
outcome of this model. The temperature of the X-ray emitting gas
is simply a function of the shock velocity. A constant X-ray temperature
therefore implies the wind velocity is constant within the
region considered. The analytical
superwind model of \citet{cc} shows that the wind velocity is
approximately constant in the freely expanding wind region,
outside the starburst region. Velocities only drop when the wind
passes through internal shocks, but rapidly increase again if the
gas is given a chance to expand.

The most serious potential constraint upon applications of this model to
NGC 253 is the energy requirement {\em if} the \halpha~emission
is energized by shocks. Following the arguments developed
by \citet{lehnert99}, the power dissipated in a shock is related
to the \halpha~luminosity by $L_{\rm shock} \approx 100 L_{H\beta} 
\approx 33 L_{H\alpha}$, based on the calculations of
\citet{binette_shocks}. This implies shocks are dissipating
the energy in the northern halo of the
superwind at a rate $L_{\rm shock} \sim 1.5 
\times 10^{40} \ergps$. Using the {\em lowest} estimate of
the nuclear supernova rate in NGC 253 of $\sim 0.03 \pyr$
from \citet{mattila2001}
(\citet{ua97} provide an upper limit of  $< 0.3 \pyr$), 
the mechanical power
flowing into the northern halo in the superwind is
$\sim 5 \times 10^{41} \eta_{\rm therm} \ergps$, based on the
models of \citet{lh95}. Even if we
adopt a low value of the supernova energy thermalization
efficiency, \eg $\eta_{\rm therm}$ = 10\%, supernovae can provide
more than enough energy to power the \halpha~emission through shocks. 

Any contribution
to the \halpha~emission from photoionization by the starburst
will reduce the implied energy drain on the superwind.
Using the arguments of \citet{lehnert99}, we can compare the
number of ionizing photons required to produce the halo \halpha~emission
with the number of emitted in the direction of the northern halo by
that starburst, and see if the required ``escape fraction'' is reasonable.
The \halpha~luminosity of the northern halo requires 
$N_{\rm Lyc} \sim 4 \times 10^{50} \ps$ to maintain. The nuclear starburst
produces $N_{\rm Lyc} > 10^{53} \ps$ \citep{engelbracht1998}, 
of which a fraction $\Omega/4\pi \sim$
0.1 -- 0.2 travel in the direction of the northern halo.
Hence only $\lesssim 4$\% of these ionizing photons need to escape
into the halo to explain the observed \halpha~emission by 
photoionization alone. This is consistent with current estimates of UV photon
escape fractions (\eg \citet{leitherer_uvescape}; 
\citet{heckman_uvescape}) in star-forming galaxies.

Is there any evidence for clouds of neutral gas, or tidal
debris, in the halo of NGC 253? 
The H{\sc i} studies of \citet{puche91} and \citet{koribalski95} 
show no H{\sc i} structures outside the disk,
placing an upper limit on the hydrogen column density in the
halo of $> 2.4 \times 10^{20} \pcmsq$ within a $\sim 1\arcmin$ beam.
The tidal H{\sc i} features in the halo of M82 have column
densities of $\lesssim 1.5 \times 10^{20} \pcmsq$ \citep{yun93}, so
similar features could well be present in the halo
of NGC 253 and have escaped detection in the existing H{\sc i}
surveys.

Unless clouds that are over-run by the wind are destroyed on  time-scales
short compared to the age of the wind, it is difficult to 
explain the roughly X-shaped morphology of the \halpha~and 
soft X-ray emission in NGC 253 with this model. 
 
As halo ``clouds'' or tidal debris 
are not necessarily symmetrically distributed
around the galaxy, this model provides a simple explanation of the
often-mentioned assymetries in superwinds like M82 and NGC 253
(\eg see discussion
in \citet{shopbell98}), where one pole is typically
brighter and more extended that the opposite pole of the outflow.
Nevertheless, it remains to be seen how significant wind assymetries
are in larger samples of starburst-driven winds.

\subsubsubsection{Model 3: Shocked disk gas carried into the halo}

Motivated by the observations of outflowing \halpha-emitting gas
in the inner several kpc of superwinds, the closely associated
soft X-ray emission, and numerical simulations (in particular
those of \citet{suchkov94}), an alternative model
is that the superwind drives cool ambient gas from the disk into
the halo, where it is responsible for the observed \halpha~emission.
The X-ray emission arises in the vicinity of this cool gas, either
in stand-off bow shocks in the superwind material itself, or perhaps
in conductive or turbulent mixing interfaces around the cool
clumps (see Fig.~\ref{fig:cartoon}b).

The recent {\it ISOPHOT} observations of NGC 253 by \citet{radovich_isophot},
which show X-shaped FIR emission from dust in the same location as
the X-ray and \halpha~emission are most easily explained by this model,
as the dust is dragged up from the disk along with the \halpha~emitting 
gas (see also \citet{phillips93}; \citet{alton99}; \citet{heckman2000}).

Most of the arguments presented in Model 2 can be directly
transferred to this model, the only real difference is that the wind
brings its own cool gas along with it instead of requiring
a pre-existing population of halo clouds. The shock energy
dissipation and photoionization arguments about the
\halpha~emitting gas remain the same. The temperature of the
X-ray emitting gas may again be  related to the wind velocity,
although it may be more of an emission-weighted mean gas temperature
if conductive interfaces or mixing layers dominate in X-ray emissivity over 
shocked wind.

Simulations show that cool dense gas from the disk of the galaxy
is transported into the halo along the walls of the cavity, as
Kelvin-Helmholtz instabilities rip off clumps of gas and the wind
accelerates them into the halo (see Figs.~3 \& 6 in \citet{suchkov94},
and Fig.~11 of \citet{heckman2000}). This naturally leads to an
X-shaped \halpha~morphology extending out of the disk and into the halo. 
A thin X-ray-emitting skin covers the
clumps of cool gas dragged into the halo. Note that these structures are well
within the confines of the wind -- they do not mark the outer shock
propagating into the halo. 

The general \halpha~and X-ray morphology
predicted in this model is very similar to that we observe in NGC 253.
One minor difference between this model and the observations is that
it does not directly predict
the systematic offset of the X-ray emission
with respect to the \halpha~emission that may exist in the
northern halo of NGC 253 (Fig.~\ref{fig:halpha_zoom}). 
If the X-ray emission is systematically offset
with respect to the interior of the \halpha~emission then Model 3 is
in trouble.

\subsubsubsection{Model 4: A hot swept-up shell of halo gas}

A useful theoretical approximation is to consider superwinds
as superbubbles (\citet{weaver77}; \citet{maclow_mccray88})
expanding in a low density halo medium, \ie to
ignore the presence of the denser ISM in the disk once the superbubble
has blown out into the halo.

A superbubble can be described as a set of concentric regions. Proceeding
from the center outwards these are: A central region of mass and energy
injection (the starburst region). Out of the starburst
region flows a  supersonic wind of thermalized
SN-ejecta and stellar wind material. 
The gas properties in these two regions are well-described 
by the analytical model of \citet{cc}. This region of freely expanding
wind is bounded by a shock, commonly referred to as the reverse
shock, which separates the free wind from a region of hot shock-heated
wind material. A contact discontinuity separates the shocked-wind
from a surrounding shell of swept-up shocked ambient gas. An outer
shock marks the boundary between the swept-up shell and the surrounding
undisturbed ambient medium.

If the cooling time of the shocked ambient gas is small compared to the
the age of the bubble, the shell cools to $T \lesssim 10^{4}$ K and
forms a thin dense radiative
shell,  surrounding the non-radiative hot shocked-wind interior to
it. This is the physical situation most commonly invoked when considering
wind-blown bubbles or superbubbles. Given the high mechanical
energy injection rates in starbursts, along with the assumption
that the gas density in the halo is low, it is quite possible that
the swept-up shell may still be hot and an appreciable source
of soft X-ray emission (Fig.~\ref{fig:cartoon}c). 
We will discuss the case of a cool shell
in Model 5 below.

Is this model consistent with the observational data?
For a bubble driven into a medium of
constant density $\rho_{0}$, by a wind with a mechanical power 
$\dot E_{\rm w} = L_{\rm w}$, the radius of the bubble
as a function of time $t$
is $R_{s} = 0.76 L_{\rm w}^{1/5} \, \rho_{0}^{-1/5} \, t^{3/5}$
\citep{castor75}. The age of the bubble, its radius and the
velocity of the outer shock are related by $t = 0.6 R_{\rm s} / v_{\rm s}$.
Is we assume that the velocity of the outer shock in the halo of
NGC 253 is $v_{\rm s} = 470$ km/s (which gives a post-shock temperature of
$3 \times 10^{6}$ K), and the radius $R_{\rm s} \sim 10$ kpc, then
the dynamical age is $t_{\rm dyn} = 12.5$ Myr. 
Adopting the lower estimate of the SN rate discussed in Model 2,
and assuming the thermalization efficiency is unity, $L_{\rm w} \sim
10^{42} \ergps$. Solving for $\rho_{0}$ we obtain 
$\rho_{0} = 5.5\times10^{-28}$ g$\pcc$, implying a total initial
halo mass of $M \sim 4\pi \, R_{\rm s}^{3} \rho_{0} \sim 3.4 \times 10^{7}
\Msol$ of gas within a radius $R_{\rm s}$ has been swept-up and shock-heated.

The cooling time of this gas is longer than the dynamical age of the bubble,
as required to maintain a hot shell.
For isobaric cooling at Solar abundance, the cooling time $t_{\rm c}$ 
from an initial temperature of $3\times10^{6}$ K down to $10^{4}$ K
is $t_{\rm c} = 2.8\times10^{5} n_{\rm e}^{-1}$ years, 
based on numerically integrating the non-ionization 
equilibrium cooling curves of 
\citet{sutherland_dopita93}\footnote{A useful approximation based on
these calculations is that for isobaric cooling at Solar abundance
$t_{\rm c} n_{\rm e} \approx 4000 \, \yr \, \pcc$  for $4.4 < \log T < 5.4$,
and $t_{\rm c} n_{\rm e} \approx 4000 \, (T/10^{5.4})^{1.7} \, \yr \, \pcc$ 
for $5.4 \le \log T < 8.0$.}. Assuming the outer shock is strong,
the electron number density in the shocked shell of halo gas is
$n_{\rm e} \sim 2 \rho_{0} / \mu \sim 1.1 \times 10^{-3} \pcc$, 
where $\mu = 10^{-24}$ g.
The cooling time of the halo in this model is $t_{\rm c} \sim 2.5 \times
10^{8}$ years, an order of magnitude greater than the dynamical
age of the bubble. 

The implied X-ray emission integral $EI = n_{\rm e} n_{\rm H} V
\approx n_{\rm e}^{2} \times M / 2\mu n_{\rm e} \sim 3.8 \times 10^{61} \pcc$,
agreeably close to the order-of-magnitude observation estimates
of the emission integral presented in \S~\ref{sec:spectra:properties}.

Thick, X-ray-emitting shells of swept-up shocked halo gas
are found in the numerical simulations of 
both \citet{suchkov94} and \citet{ss2000}, where the
mean expansion velocities of the wind are $\sim 1000$ km/s in the halo
and densities are too low to allow significant cooling.
The 2-D X-ray emissivity maps in \citet{suchkov94} demonstrate that
the largest volume of moderately high X-ray emissivity gas in
0.1 -- 2.2 keV energy band is this shell of shocked halo gas,
and it contributes approximately half of the total soft X-ray luminosity.

This does not prove that the X-ray emitting gas in the
halo of NGC 253 is a hot shell of swept-up shell of halo gas,
merely that this model is consistent with the X-ray data.

The most significant and troubling problem with this model is 
where does the spatial correlation with the
\halpha~emission come from? The most plausible origin for X-shaped
distributions of cool gas in the haloes of starburst galaxies
is that it is dragged up by the wind, from along the walls of the
superwind cavity within the disk, \ie as presented in Model 3.
The cartoon of Model 4 presented in Fig.~\ref{fig:cartoon}c,
which is based on numerical simulations, would
imply that the \halpha~emission would lie within the X-ray emission,
which does not seem to be the case.

If this model (or Model 3, which also predicts hot gas beyond the
region of dense \halpha~emitting gas) is correct, the 
{\it XMM-Newton} observations
of NGC 253 might be able to detect the faint presence of the
wind beyond the \halpha~emitting regions, given the larger field of
view of the EPIC instruments compared to ACIS.
For other more distant edge-on starbursts like NGC 3628 and NGC 3079
the ACIS field of view is large enough to test the idea that X-ray
emission may extend beyond any \halpha~emission. 

\subsubsubsection{Model 5: Swept-up thick-disk and shocked wind}

If the swept-up shell does cool, the result is a \halpha-emitting
shell surrounding a hot bubble of shocked SN-ejecta. In the
\citet{weaver77} wind-blown bubble model, X-ray emission from this 
region of hot, shocked, gas
is concentrated near the cool shell, because thermal conduction
partially evaporates the shell into the hot bubble, locally
increasing the X-ray emissivity.

This naturally produces limb-brightened \halpha~emission with
soft X-ray emission concentrated slightly interior to the
\halpha~emission. Applied to superwinds expanding into a galactic
halo, the observed morphology would approximate a figure-eight
centred on the starburst nucleus if the gas density in the halo is uniform,
or an open-ended X is the halo gas density decreases with increasing
height above the plane of the galaxy (as depicted in Fig.\ref{fig:cartoon}d).

Given the lack of an obvious cap
to the superwind in NGC 253, a plausible model is that this 
halo material represents
some form of thick disk component with a scale-height of several kpc, 
possibly lifted above the thin disk 
by the star-formation activity within the disk, as
seen in the numerical simulations of \citet{rosen_bregman}.
Narrow-band optical imaging strongly suggests that the presence
of extended \halpha~emitting gas in the halos of non-starburst
galaxies correlates with the strength of star-formation activity
in the disk (see for example 
\citet{rand96}; \citet{wang97}; \citet{hoopes99}).
If so, then the expansion of the superwind
into this material can be used as a probe of the distribution
of gas in the halos of disk galaxies \citep{sofue_vogler}.

In order for the shocked halo to cool, the shock velocity has to be relatively
low and the post-shock density reasonably high -- effectively the more
gas there is in the halo the more likely it will be for the swept-up
shocked halo to cool rapidly. For convenience we assume a shock velocity of
210 km/s, as the resulting post-shock temperature is $6 \times 10^{5}$ K.
This is the critical temperature adopted by \citet{castor75} below
which they assumed the shell cools rapidly.

Following the same method as presented in Model 4, the dynamical age
of the bubble is $27.9$ Myr, and for $L_{\rm w} = 10^{42} \ergps$ the
required initial density and gas mass are 
$\rho_{0} = 6.2 \times 10^{-27}$ g$\pcc$ and $M = 3.8 \times 10^{8} \Msol$.

The initial-post shock electron number density, prior to cooling, is 
$n_{\rm e} = 1.24 \times 10^{-2} \pcc$. The isobaric cooling time
at $T = 6 \times 10^{5}$ K is $t_{\rm c} 
\approx 1.8 \times 10^{4} \, n_{\rm e}^{-1}$, \ie 
$ \approx 1.5 \times 10^{6}$ years.
As this is $\sim 1/20$th of the dynamical age of the wind,
it is clear that the shocked halo can cool down to $T \sim 10^{4}$ K,
and be a source of the observed \halpha~emission. Note that the 
total mass of swept-up gas in the example above is larger than the 
observed mass of \halpha~emitting gas in the northern halo, which we estimate
to be $M_{\rm H\alpha} \approx 4 \times 10^{6} \eta_{\rm H\alpha} \Msol$
(where $\eta_{\rm H\alpha}$ is the fraction of the total halo volume
currently occupied by the \halpha~emitting gas). In principle there
is nothing preventing the dense gas in the shell cooling below $10^{4}$
K and hence not contributing to the observed \halpha~flux, 
so the requirement that there is a few times $10^{8} \Msol$ of
gas in the halo is not a significant problem for this model.
The average hydrogen column density through the halo assuming a
total mass of $3.8 \times 10^{8} \Msol$ is crudely 
$\nH \sim M/4\pi R^{2}_{\rm s} \sim 1.3 \times 10^{20} \pcmsq$,
below the current observation limits for \hi~studies of
NGC 253.

An interesting prediction of this model is a high O{\sc vi} flux
from the cooling region behind the outer shock (provided 
$v_{\rm s} \gtrsim 150$ km/s). For the example parameters
discussed above, we predict a O{\sc vi} column density through the
shock of $5\times 10^{14} \pcmsq$, and a total emitted luminosity 
in the O{\sc vi} $\lambda \lambda$ 1032, 1038 doublet of 
$\sim 7 \times 10^{40} \ergps$. A large fraction of the energy radiated
by the cooling shocked halo gas is carried  by O{\sc vi} emission.

This model has the appealing conceptual 
advantage over the previous models in that it easily
explains the morphology of the \halpha~and X-ray emitting gas, and
leads to the interesting possibility that we might probe be able to 
observationally probe some function of the halo gas density 
and superwind ram pressure at $\sim 10$ kpc distances above the
plane of the galaxy. In contrast the
\halpha/Xray morphology in Model 3 does not tell us the location of the
outer shock of the wind, instead depending on the complex physics of
gas entrainment from the disk by the wind. Nevertheless, the viability
of this model depends crucially on the ability of the shocked halo material
to cool, and this model predicts larger gas masses in the halo than
implied by the observed \halpha~flux alone.
More sensitive H{\sc i} observations of NGC 253 would be of great use
for constraining both this model and Model 2.

\subsection{Magnetic fields in the halo}

We have ignored the role of magnetic fields in the previous
discussion. This appears justified based on a comparison between the
energy density in hot gas with that in magnetic fields and
cosmic rays. The minimum energy B-field strength in the halo
is $\sim 2 \, \mu$G (\citet{carilli_n253radio}; \citet{beck1994}).
The ratio of the energy density in this field to the thermal
pressure of the X-ray emitting plasma is $\sim 0.025 \eta_{\rm X}^{0.5}
{\cal R}^{-0.5}$, which suggests magnetic fields are dynamically unimportant.

The orientation of the magnetic field in the lower halo of NGC 253 
($z \sim 3$ kpc, \ie
roughly at the base of the northern ridge) is parallel to the disk
\citep{beck1994},
in contrast to the radial magnetic field structure in the halos
of M82 \citep{reuter1994} and NGC 4631 \citep{golla_hummel94}.
This reasonably lead \citet{beck1994} to suggest that the magnetic field
structure in halo of NGC 253 suppressed the superwind from flowing further
into the halo. This now seems unlikely, given the relatively weak field
estimated above. Further evidence against the idea of wind suppression, 
and supporting
the idea that magnetic fields are dynamically weak, is
the clear presence of both X-ray and \halpha~emission further
out in the halo beyond the region of parallel B-fields..

\subsection{Implications for understanding the superwind phenomenon}

It is worthwhile to take a step back from the minutiae of the different
models presented in the previous sections to consider the broader 
implications of these observations for understanding starburst-driven 
outflows.

Strong similarities between \halpha~and X-ray morphology
over a range of spatial scales, from 100's of pc to $\sim 10$ kpc, 
have been found in both M82's starburst-driven superwind 
(see in particular \citet{watson84}; \citet{dwh98};
\citet{shopbell98}; \citet{devine99} \& \citet{lehnert99})
and the ultra-luminous IR galaxy Arp 220 (\citet{heckman_arp220};
Jonathan McDowell, private communication).
In the other nearby edge-on starbursts with superwinds, such
as NGC~253, NGC~3079 \& NGC~3628 such global (10 kpc-scale) 
\halpha/X-ray relationships had not previously been found
(\citet{dahlem_n3628}; \citet{rps97};
DWH98; \citet{pietsch_n3079}; \citet{pietsch2000}). 
The detection of \halpha~emission similar in flux the X-ray emission
in the halo of NGC 253 (a far more typical starburst galaxy
than M82 or Arp 220) suggests that {\em all} superwinds may have
large-scale \halpha~halos.

Physically meaningful models of superwinds {\em must}
 be able to simultaneously
explain the presence, relative spatial distribution and 
relative fluxes of X-ray and
\halpha~emission in the haloes of starburst galaxies. Models
that satisfy X-ray constraints only, for example delayed recombination
\citep{breitschwerdt99} or the Bipolar Hypershell model of
\citet{sofue_vogler}, fail to predict the observed \halpha~emission.

Our ability to infer the properties of superwinds depends on knowing
the physical origin of the emission. As the previous section has
shown, there are a variety of possible models that can explain
spatially correlated X-ray and \halpha~emission. Depending on
which model is chosen different inferences about superwind
properties can be drawn. For example, in Model 2 the X-ray-emitting
plasma is shocked superwind and its temperature a measure of the
wind velocity, while in Model 3 the X-ray emission may come
from conductive interfaces or turbulent mixing layers
around clouds, and its  temperature may have little to do with
the wind velocity. Instead the X-ray and \halpha~emission tell us
about mass entrainment of disk ISM by the wind and its transport
into the halo.
In Model 5 the presence of \halpha~emission 
immediately places limits on the velocity of the outer shock
and the properties of pre-existing halo gas, whereas it
is the X-ray data that encodes this information in Model 4.
In models 4 and 5 the X-ray and \halpha~emission mark the
outer boundary of the wind, in Model 3 the wind can extend
far beyond the \halpha~and X-ray bright region.
All of the above models have been selected in an attempt to 
satisfy the (admittedly loose) observational constraints,
yet they each carry very different implications for our
understanding of winds.

If the \halpha~emission is due to cool gas transported out of the disk
by the action of the superwind (Models 3 \& 4), it suggests that
it is not only the hottest gas phases that escape the gravitational
potential of the host galaxy. 

Superwinds are without-doubt extremely complex, and it is
quite possible that two or more  of
the different models presented above
may apply within a single galaxy. The best interpretation for the
X-ray/\halpha~ridge 12 kpc north of M82 is the halo cloud
model of \citet{lehnert99}, but Models 3, 4 or 5 may best
describe the emission closer to M82's disk.

The most fundamental discriminant between these models
are the details of the exact spatial location of the \halpha~emission
with respect to the X-ray emission. More quantitative properties
are unfortunately biased by the long standing problems
of observationally unknown volume filling factors and the
systematic problems associated with the X-ray-determined abundance
determinations. In NGC 253 the limited field of view of the ACIS
instrument has somewhat restricted our ability to apply these
discriminants, but for slightly more distant starbursts, \eg NGC 3628
and NGC 3079, {\it Chandra} will be the ideal instrument.

\section{Summary}
\label{sec:summary}

We have presented a detailed case study of the diffuse X-ray emission
from a starburst-driven superwind in the closest typical starburst galaxy,
NGC 253. The main results of this study can be summarized as:

\begin{enumerate}
\item The diffuse X-ray emission appears to outline an X-shaped structure
  centered on the starburst nucleus, although the north western and south
  western arms of the X are relatively indistinct. The ridge-or-arc-like
  feature in the northern halo forms part of the bright north eastern
  limb of the X. 
\item We have uncovered a similar, previously unreported,
   X-shaped-feature in \halpha~images
  of NGC 253, very similar to the soft X-ray emission and extending
  $\sim 8$ kpc above the plane of the galaxy. Very extended \halpha~emission
  is likely to be a general feature of all superwinds.
\item Based on our {\it Chandra} data, 
  there is no statistically significant variation in the spectral
  properties of the diffuse X-ray emission in the northern halo, over scales
  of several 100 pc to several kpc.
\item Spectral features due to highly ionized O, Fe, Mg and possibly other
  elements robustly demonstrate a thermal origin for at least 
  some fraction of the
  soft X-ray emission in the halo of NGC 253, 
  although the unphysically low abundances
  derived from spectral fitting indicate that we do not fully understand
  the X-ray spectra of starbursts. One possibility is a non-thermal
  X-ray contribution, perhaps synchrotron emission associated with the
  cosmic rays and magnetic fields dragged into the halo by the superwind.
\item  The marginal detection
  of diffuse emission in the 2 -- 8 keV band is most likely due to
  background  subtraction problems, and not due to a very hot superwind
  component. We can only place upper 
  limits on the presence of any spectrally-hard
  diffuse component, with $L_{\rm X,hard} <  10^{38} \ergps$ assuming
  a plasma temperature of $kT = 4$ keV.
\item Soft X-ray and \halpha~fluxes are very similar in the northern halo,
  consistent with the X-ray to \halpha~flux ratios found in
  other starbursts, and to those in NGC 253's  nuclear outflow cone.
\item The total X-ray and \halpha~luminosities of the northern halo
  region are $L_{\rm X} \sim  4.9 \times 10^{38} \ergps$ (0.3 -- 2.5 keV) and 
  $L_{\rm H\alpha} \sim 4.4 \times 10^{38} \ergps$. 
  Other plasma properties are more uncertain,
  given the unknown metal abundance and filling factor of the
  X-ray-emitting plasma. Photoionization by UV photons escaping the
  starburst can explain the ionization of the \halpha~emitting gas.
  Alternatively, if 
  the \halpha~emission were powered by shocks alone this implies a
  power dissipation in the northern halo
  shocks of order $\sim 1.5\times10^{40}\ergps$. This is value is still
  well below the minimum energy injection rate by the starburst, so
  we conclude radiative energy losses in superwinds are minimal.
\item The spatial distribution of the \halpha~and {\it Chandra}
  X-ray emission in the halo 
  is similar on several kpc scales, but at smaller scales
  there is no clear correlation or anti-correlation. In the northern
  halo the X-ray emission appears offset towards the interior of the superwind
  from the \halpha. In the southern halo the X-ray and \halpha~emission
  do not appear to be offset from each other, although this
  inference is based on the lower spatial resolution {\it ROSAT} PSPC data.
  Successful models of superwinds must be able to simultaneously explain
  both the X-ray and \halpha~emission.
\item \citet{sofue_vogler} have pointed out that the location
  of the outer boundary of the
  X-ray emission in superwinds may provide a convenient method of probing
  the relatively unknown density distribution in the halo of galaxies.
  This relies upon the observed X-ray emission marking the edge of the wind,
  which we emphasize is not necessarily the case. Explaining the
  \halpha~emission  we have found, which appears to envelope the 
  X-ray emission in the halo NGC 253, in a Sofue \& Vogler-like model
  appears difficult. Models based on the \citet{weaver77} wind-blown
  model show that the shell of swept-up halo gas may be the origin of the
  \halpha~emission in the halo, if there was a sufficient density
  of gas in the halo to allow it to cool in less than a dynamical time.
\item We present and discuss a variety of
  plausible models that aim to explain the correlated X-ray and
  \halpha~emission. There are a variety of
  possible physical origins of the X-ray and \halpha~emission, each which
  have different implications for our understanding of the properties of
  superwinds and the halo gas they expand into.
  We find that the relative
  spatial location of the \halpha~emission with respect to the X-ray
  emission is the most robust discriminant between the various models.
\end{enumerate}

In the longer term a 
combination of this detailed study of the halo of the nearest
classic starburst NGC 253, combined
with {\it Chandra} and \halpha~observations
of a larger sample of superwinds,
will determine which of the possible modes of coupled X-ray and 
\halpha~emission is the most common in superwinds.

\acknowledgments

Is is a pleasure to thank the anonymous referee 
for a considered, helpful, and prompt report.
We would like to thank Jonathan McDowell, Tahir Yaqoob, 
Ed Colbert, Nancy Levenson \& Nichole King
for helpful advice and suggestions.
The team that built and operates {\it Chandra} also deserves
recognition, for their hard work, and for making such an excellent
telescope.
DKS is supported by NASA through {\it Chandra} Postdoctoral Fellowship Award
Number PF0-10012, issued by the {\it Chandra} X-ray Observatory Center,
which is operated by the Smithsonian Astrophysical Observatory for and on
behalf of NASA under contract NAS8-39073.
This research has made use of data obtained through 
the High Energy Astrophysics Science Archive Research 
Center Online Service, provided by the NASA/Goddard Space Flight Center.
This work was supported in part by NASA through 
grants LTSA NAG56400 and GO0-1008X.

\appendix

\section{Source searching methods}
\label{sec:appendix:source_searching}

We used two different techniques to search for X-ray point sources in the halo
of NGC 253: the wavelet-based
{\sc Wavdetect} and the simpler sliding box method {\sc Celldetect}.
Both of these are routines provided as part of {\sc Ciao}. 
Each of these methods
has its own distinct set of strengths and weaknesses.
Comparing sources detected by both methods to the aperture method
described below revealed biases in both the wavelet and sliding box
source detection algorithms. 
{\sc Wavdetect} provided apparently accurate estimates of the total
number of source counts in each source, but 
the uncertainties on the total source counts it quotes were sub-Poissonian and
hence overestimated the significance of the faintest sources.
{\sc Celldetect}, on the other hand, gave realistic uncertainties
for a given number of source counts, but systematically underestimated
the total counts in each source and hence underestimated the source
significance.

Both of these methods detected large numbers of apparently faint
point sources that appear to be spurious mis-identifications of
local enhancements in the diffuse emission as point sources,
given that each method detected a different set of sources at
low $S/N$, that these sources appeared to follow the distribution of
the diffuse emission, and that these detections were in the
soft 0.3--2.0 keV X-ray band.

Source searching the halo region of NGC 253 is relatively difficult,
given the presence of both moderate and low surface brightness diffuse
emission, irregularly distributed over the chip.
The biases in the
standard source detection algorithms
may well be related to the unavoidable difficulty of robustly
detecting point sources under such conditions.
We therefore used both {\sc Wavdetect} and {\sc Celldetect} 
to identify point {\em source candidates}. To
find the  true count rate, count rate uncertainties and
source significance of any candidate, we used a method
inspired by the source
searching procedures described in \citet{giaconni2001}.

A list of candidate sources was created by source searching
images of the entire S3 chip created in three energy bands
(soft X-rays: 0.3--2.0 keV, hard X-rays: 2.0-8.0 keV and
broad band: 0.3-8.0 keV energy band) with both wavelet and
sliding box methods. All six resulting lists of candidate sources were
merged, accepting as a potential source any detection
with a signal-to-noise ratio (S/N) greater than 1.5 using either
method, but removing any duplicate detections.

We then calculated the source count rate and uncertainty
by measuring the counts in a circular source region, and in
an encompassing local background region. To estimate the
true source counts and uncertainty from these measurements
we used the method detailed in \S~5 of the {\sc Ciao} Detect
Guide \citep{detect_guide}, which uses Gerhel's (1986) formula
for the uncertainty
($\sigma_{N} \approx 1 + \surd{(0.75 + N)}$) as a good 
approximation to the Poissonian nature of the errors.
Note that this is the same theory as is used in
the {\sc Celldetect} source searching algorithm. The main
differences between our method and {\sc Celldetect} are:
(a) the source and background regions we use are
larger than those used by {\sc celldetect} (see below), and (b) we ensure
that no background or source region encloses counts from any
other nearby point source (a problem that will have affected
{\sc Celldetect} in several instances for this observation).

We empirically derived the variation in the size of the ACIS PSF
as a function of off-axis angle in the S3 chip using 12 bright
point sources, using this observation  and our separate
S3 chip observation of the disk of NGC 253, 
by fitting two dimensional Gaussian models to
each point source. This allows us to scale the size of the source
extraction region used to calculate the source counts rates
 realistically with off-axis angle. We used a source
extraction radius equal to the maximum of either $3\arcsec$ or 
the 4 sigma radius of the empirical PSF (the $4\sigma$ radius is 1\farcs35
on-axis, and 9\farcs4 at an off-axis angle of $6\arcmin$). 
We used a circular background region of radius
$15\arcsec$ in all cases, which always fully enclosed the source region, but
in many cases was offset from the center of the source region to avoid
including counts from other nearby source candidates.
These source and background regions are significantly larger than
those used in {\sc Celldetect} (for example, the on-axis source region
is a square region of  only 9 pixels, whereas our smallest
source region comprises $\sim 110$ pixels). As significant numbers
of counts can be seen to fall outside the standard 9 pixel {\sc Celldetect}
boxes in the brighter sources, we believe that this leads to {\sc Celldetect}
underestimating the total source count rates. Experimentation showed
that as the size of the source and background regions are increased (starting
from the same size as used by {\sc Celldetect}), 
the estimated source count rates increased.
Once the size of the source and background regions 
approached the values we adopted, the estimated source counts and errors
had converged.

Of an initial list of 86 candidate sources, this method arrived at a 
total of 30 unique sources of $S/N \gtrsim 2.0$, of which 13 sources
had $S/N \gtrsim 3.0$. Although a S/N ratio of 2.0 is low, many of
these low significance sources are likely to be genuine, as they
have clear optical counterparts in deep CCD images 
(see \S~\ref{sec:data:astrometry}).
The positions, count rates, fluxes and any optical identifications
of these sources are given in Table~\ref{tab:point_sources}.
The brightest of the sources, \eg source number 20, may be of use in
the future for UV and X-ray absorption line studies of superwinds
(\eg \citet{norman_abs}).

\begin{deluxetable}{lrrrcccrl}
 \tabletypesize{\scriptsize}%
\tablecolumns{9} 
\tablewidth{0pc} 
\tablecaption{X-ray point sources in the northern halo
	\label{tab:point_sources}} 
\tablehead{ 

\colhead{ID} & \colhead{Official name} 
	& \colhead{$\alpha$ (J2000)} & \colhead{$\delta$ (J2000)}
	& \multicolumn{3}{c}{Background-subtracted counts\tablenotemark{a}} 
	& \colhead{$f_{\rm X}$\tablenotemark{b}}
	& \colhead{Previous\tablenotemark{c}} \\
\colhead{ } & \colhead{(CXOU number)}
	& \colhead{(h m s)} & \colhead{($\degr$ $\arcmin$ $\arcsec$)}
	& \colhead{(broad band)} & \colhead{(soft band)} 
	& \colhead{(hard band)} & \colhead{ }
	& \colhead{detections} }
\startdata
1 & J004659.1-250814 & 00 46 59.06 & -25 08 14.5
	& $16.0\pm{8.8}$ & $14.8\pm{7.3}$ & $1.6\pm{5.9}$ 
	& $0.24\pm{0.13}$
	& \nodata \\
2 & J004659.7-251209 & 00 46 59.68 & -25 12 08.6
	& $17.9\pm{6.8}$ & $12.0\pm{5.4}$ & $5.9\pm{4.8}$ 
	& $0.27\pm{0.10}$
	& \nodata \\
3 & J004701.1-250843 & 00 47 01.13 & -25 08 42.8
	& $7.2\pm{9.0}$ & $-7.2\pm{6.6}$ & $14.4\pm{7.0}$ 
	& $0.11\pm{0.14}$
	& \nodata \\
4 & J004701.9-251108 & 00 47 01.89 & -25 11 08.1
	& $13.8\pm{7.2}$ & $13.3\pm{6.5}$ & $0.5\pm{4.1}$ 
	& $0.21\pm{0.11}$
	& \nodata \\
5 & J004703.0-251028 & 00 47 03.03 & -25 10 27.9
	& $13.8\pm{6.8}$ & $-0.1\pm{4.3}$ & $13.8\pm{5.9}$ 
	& $0.21\pm{0.10}$
	& \nodata \\
6 & J004708.6-250714 & 00 47 08.60 & -25 07 14.0
	& $48.2\pm{11.7}$ & $36.0\pm{8.9}$ & $12.2\pm{8.4}$ 
	& $0.73\pm{0.18}$
	& \nodata \\
7 & J004709.6-251404 & 00 47 09.61 & -25 14 04.2
	& $410.0\pm{21.4}$ & $383.3\pm{20.7}$ & $26.7\pm{6.4}$ 
	& $6.21\pm{0.32}$
	& R5, DWH3, VP13  \\
8 & J004709.8-251513 & 00 47 09.85 & -25 15 13.2
	& $15.0\pm{5.6}$ & $9.6\pm{4.6}$ & $5.4\pm{4.0}$ 
	& $0.23\pm{0.08}$
	& \nodata \\
9 & J004711.2-251333 & 00 47 11.18 & -25 13 33.4
	& $11.5\pm{5.0}$ & $11.8\pm{4.9}$ & $0.2\pm{2.3}$ 
	& $0.17\pm{0.08}$
	& \nodata \\
10 & J004712.8-251452 & 00 47 12.80 & -25 14 51.6
	& $9.8\pm{4.9}$ & $8.7\pm{4.6}$ & $1.1\pm{2.7}$ 
	& $0.15\pm{0.07}$
	& \nodata \\
11 & J004713.1-251240 & 00 47 13.14 & -25 12 39.7
	& $9.3\pm{4.6}$ & $4.4\pm{3.6}$ & $4.9\pm{3.6}$ 
	& $0.14\pm{0.07}$
	& \nodata \\
12 & J004716.1-250713 & 00 47 16.14 & -25 07 13.0
	& $35.4\pm{10.1}$ & $8.1\pm{7.1}$ & $27.3\pm{7.9}$ 
	& $0.54\pm{0.15}$
	& \nodata \\
13 & J004716.2-251306 & 00 47 16.19 & -25 13 06.3
	& $11.8\pm{5.0}$ & $8.1\pm{4.3}$ & $3.7\pm{3.4}$ 
	& $0.18\pm{0.08}$
	& \nodata \\
14 & J004717.6-250603 & 00 47 17.65 & -25 06 02.9
	& $21.8\pm{9.8}$ & $18.6\pm{8.0}$ & $3.2\pm{6.5}$ 
	& $0.33\pm{0.15}$
	& \nodata \\
15 & J004718.3-251006 & 00 47 18.33 & -25 10 05.7
	& $38.9\pm{7.7}$ & $10.8\pm{4.9}$ & $28.1\pm{6.6}$ 
	& $0.59\pm{0.12}$
	& \nodata \\
16 & J004719.8-250643 & 00 47 19.76 & -25 06 43.2
	& $118.4\pm{14.9}$ & $113.4\pm{13.9}$ & $5.0\pm{6.6}$ 
	& $1.79\pm{0.23}$
	& \nodata \\
17 & J004721.0-251003 & 00 47 20.98 & -25 10 03.1
	& $25.6\pm{6.7}$ & $13.6\pm{5.2}$ & $12.0\pm{4.9}$ 
	& $0.39\pm{0.10}$
	& \nodata \\
18 & J004722.5-251200 & 00 47 22.49 & -25 12 00.3
	& $79.8\pm{10.2}$ & $10.4\pm{4.7}$ & $69.5\pm{9.5}$ 
	& $1.21\pm{0.15}$
	& \nodata \\
19 & J004722.6-251125 & 00 47 22.57 & -25 11 25.0
	& $13.6\pm{5.2}$ & $9.9\pm{4.6}$ & $3.7\pm{3.4}$ 
	& $0.21\pm{0.08}$
	& \nodata \\
20 & J004723.1-251054 & 00 47 23.15 & -25 10 54.0
	& $439.5\pm{22.1}$ & $357.4\pm{20.0}$ & $82.0\pm{10.1}$ 
	& $6.66\pm{0.33}$
	& R9, DWH8, VP22  \\
21 & J004723.2-251122 & 00 47 23.24 & -25 11 22.4
	& $9.5\pm{4.7}$ & $2.6\pm{3.4}$ & $6.9\pm{4.0}$ 
	& $0.14\pm{0.07}$
	& \nodata \\
22 & J004723.3-250855 & 00 47 23.33 & -25 08 55.5
	& $18.3\pm{6.6}$ & $17.1\pm{6.2}$ & $1.1\pm{3.3}$ 
	& $0.28\pm{0.10}$
	& \nodata \\
23 & J004724.6-251214 & 00 47 24.55 & -25 12 13.8
	& $10.6\pm{4.9}$ & $8.0\pm{4.3}$ & $2.5\pm{3.2}$ 
	& $0.16\pm{0.07}$
	& \nodata \\
24 & J004724.6-250751 & 00 47 24.61 & -25 07 50.8
	& $16.0\pm{7.7}$ & $11.3\pm{6.2}$ & $4.7\pm{5.3}$ 
	& $0.24\pm{0.12}$
	& \nodata \\
25 & J004726.5-250832 & 00 47 26.52 & -25 08 32.3
	& $24.0\pm{7.3}$ & $15.2\pm{6.2}$ & $8.8\pm{4.7}$ 
	& $0.36\pm{0.11}$
	& \nodata \\
26 & J004727.6-251220 & 00 47 27.62 & -25 12 20.2
	& $102.4\pm{11.3}$ & $73.6\pm{9.8}$ & $28.8\pm{6.6}$ 
	& $1.55\pm{0.17}$
	& \nodata \\
27 & J004727.9-251243 & 00 47 27.88 & -25 12 43.1
	& $17.7\pm{5.7}$ & $13.1\pm{5.0}$ & $4.5\pm{3.6}$ 
	& $0.27\pm{0.09}$
	& \nodata \\
28 & J004730.8-251128 & 00 47 30.78 & -25 11 27.7
	& $20.3\pm{5.9}$ & $7.5\pm{4.1}$ & $12.8\pm{4.9}$ 
	& $0.31\pm{0.09}$
	& \nodata \\
29 & J004731.3-251127 & 00 47 31.26 & -25 11 27.3
	& $13.0\pm{5.1}$ & $5.0\pm{3.8}$ & $8.0\pm{4.1}$ 
	& $0.20\pm{0.08}$
	& \nodata \\
30 & J004731.7-251000 & 00 47 31.66 & -25 10 00.2
	& $152.3\pm{13.7}$ & $125.8\pm{12.5}$ & $26.5\pm{6.05}$ 
	& $2.31\pm{0.21}$
	& VP30  \\
\enddata 

\tablenotetext{a}{Background-subtracted number of counts in each source, in
	each of the three energy bands (broad band: 0.3 -- 8.0 keV;
	soft band: 0.3 -- 2.0 keV; hard band: 2.0 -- 8.0 keV). Count rates
	can be estimated by dividing the total counts by the exposure 
	of 39558.3 s (this does not account for the vignetting, 
	so true count rates for sources at
	large off-axis angles will in fact be marginally higher).
	Error are quoted at 68\% confidence.}
\tablenotetext{b}{Estimated source flux, in units 
	of $10^{-14} \ergps \pcmsq$ and covering the 0.3 -- 8.0 keV 
	energy band, assuming the power law spectral fit to 
	brighter X-ray sources (discussed
	in \S~\ref{sec:halo_point_sources}) 
	applies to all the sources. Assuming the same spectral
	model, 1 count in the soft band corresponds to a flux of
	$7.0 \times 10^{-17} \ergps \pcmsq$ (0.3 -- 2.0 keV energy band),
	and 1 count in the hard band to 
	$3.9 \times 10^{-16} \ergps \pcmsq$ (2.0 -- 8.0 keV energy band).
        The quoted error is based on the uncertainty in source count rate,
        as this uncertainty dominates over the uncertainty associated
        with the flux estimated from the spectral model.}
\tablenotetext{c}{Previous detections of these sources, based 
	on {\it ROSAT} PSPC and
	HRI observations. The letters refer to the paper 
	(R: Read \etal (1997), DWH: Dahlem \etal (1998), VP: 
	\citet{vp99}) and the following
	number is the source number given in that particular paper.}

\end{deluxetable}

\begin{deluxetable}{llcl}
 \tabletypesize{\scriptsize}%
\tablecolumns{4} 
\tablewidth{0pc} 
\tablecaption{Optical counterparts to X-ray sources in the northern halo
        \label{tab:optical_srcs}} 
\tablehead{ 

\colhead{ID} 
        & \colhead{$R$\tablenotemark{a}} 
        & \colhead{$\log (f_{\rm X}/f_{R})$\tablenotemark{b}}
        & \colhead{Notes} \\
\colhead{ } 
        & \colhead{(mag)}
        & \colhead{ }
        & \colhead{} }
\startdata
7 
        & 21.1 & 0.16 & \nodata \\
9 
        & 14.5 & -4.05 & Star \\
11 
        & 22.6 & -0.90 & \nodata \\
16 
        & 14.2 & -3.14 & Star \\
18 
        & 22.8 & 0.10 & \nodata \\
19 
        & 22.8 & -0.63 & \nodata \\
20 
        & 19.4 & -0.52 & AGN\tablenotemark{c}, $z=1.25$ \\
22 
        & 19.6 & -1.78 & \nodata \\
26 
        & 21.8 & -0.16 & \nodata \\
28 
        & 22.5 & -0.60 & \nodata \\
29 
        & 21.4 & -1.22 & \nodata \\
30 
        & 22.4 & 0.26 & \nodata  \\
\enddata 
\tablenotetext{a}{R-band magnitude of the optical candidate 
        for each X-ray source (see \S~\ref{sec:data:astrometry} for details).}
\tablenotetext{b}{The logarithm of the 0.3 -- 8.0 kev 
        X-ray flux to optical flux ratio 
        for each X-ray source with an optical counterpart.}
\tablenotetext{c}{Redshift from \citet{vp99}.}

\end{deluxetable}

\section{The significance of diffuse emission features}

For the purposes of completeness we describe the method used
to calculate the relationship between X-ray surface brightness
and the significance of that feature above the background.

After the removal of point sources, the remaining X-ray
emission consists of a diffuse component (of surface brightness
$\Sigma_{D}$ in units of count s$^{-1}$ arcsec$^{-2}$) 
and a background component (the soft and
hard X-ray backgrounds, with surface brightness $\Sigma_{BG}$). 
In an observation of length $t_{1}$ the total number of counts
$N_{T}$, detected within an aperture of area $A$ square arcseconds, is
the sum of the diffuse and background components $N_{T} = N_{D} + N_{BG}$.
The signal-to-noise ratio $R$ of a feature in a background-subtracted
image is
\begin{equation}
R = \frac{N_{D}}{\surd (\sigma_{D}^{2} + \sigma_{BG}^{2} 
        + \sigma_{BG}^{\prime \, 2})}
        = \frac{ \Sigma_{D} \, t_{1} \, A}
        {\surd(\Sigma_{D}  t_{1}  A + \Sigma_{BG}  t_{1} A [1 + t_{1}/t_{2}])},
\end{equation}
where $\sigma^{2}_{D} + \sigma^{2}_{BG} = \sigma^{2}_{T}$ is the 
variance in the total number of counts detected within the
aperture, and $\sigma^{\prime \, 2}_{BG}$ is the variance associated
with background image used in background subtraction, correctly
scaled to the exposure time of the observation. We use the
composite background fields provided by the CXC to estimate
$\Sigma_{BG}$, where the length of the background field observation
$t_{2} = 114866$ s for the S3 chip and 201452 s for the S2 chip.

Solving for $\Sigma_{D}$ we find
\begin{equation}
\Sigma_{D} = \frac{R^{2} \, A \pm{(A^{2} R^{4} + 4 \Sigma_{BG} R^{2} 
	A^{3} t_{1} [1 + t_{1}/t_{2}])^{1/2}}}
	{2 \, A^{2} \, t_{1}}.
\end{equation}

From the composite background fields the mean background surface
brightness in the S3 chip is $\Sigma_{BG} = 8.415\times10^{-7}$ counts s$^{-1}$
arcsec$^{-2}$ in the soft 0.3 -- 2.0 keV band, and 
$\Sigma_{BG} = 1.121\times10^{-6}$ counts s$^{-1}$
arcsec$^{-2}$ in the hard 2.0 -- 8.0 keV band.
For the S2 chip the equivalent background surface brightnesses
are $3.814 \times 10^{-7}$ and $5.892 \times 10^{-7}$ counts s$^{-1}$
arcsec$^{-2}$.
The X-ray surface brightness required to obtain a given signal-to-noise
ratio in an aperture of a given diameter for the S3 chip is given in 
Table~\ref{tab:signif_surf}.

\begin{deluxetable}{rrrrrr}
 \tabletypesize{\scriptsize}%
\tablecolumns{6} 
\tablewidth{0pc} 
\tablecaption{Diffuse X-ray surface brightness signal-to-noise ratios
        \label{tab:signif_surf}} 
\tablehead{ 

\colhead{Aperture} 
        & \multicolumn{5}{c}{$10^{6} \times \Sigma_{D}$\tablenotemark{a}} \\
\colhead{diameter} 
        & \multicolumn{5}{c}{Signal-to-noise ratio} \\
\colhead{($\arcsec$)} 
        & \colhead{$2$}
        & \colhead{$2.5$}
        & \colhead{$3$}
	& \colhead{$4$}
	& \colhead{$5$} }
\startdata
\cutinhead{Soft 0.3--2.0 keV band}
 2.5 & 21.62 & 33.22 & 47.38 & 83.49 & 129.84 \\
 5.0 &  6.06 &  9.00 & 12.57 & 21.67 &  33.27 \\
10.0 &  1.98 &  2.78 &  3.73 &  6.05 &   8.99 \\
20.0 &  0.77 &  1.03 &  1.31 &  1.98 &   2.78 \\ 
\cutinhead{Hard 2.0--8.0 keV band}
10.0 &  2.65 &  3.72 &  4.99 &  8.11 & 12.07 \\
20.0 &  1.01 &  1.35 &  1.73 &  2.65 &  3.73 \\ 
\enddata 
\tablenotetext{a}{X-ray surface brightness in units of
	counts s$^{-1}$ arcsec$^{-2}$, assuming an exposure time
	of 39588.3 s in the soft band and 29439.6 s in the hard
	X-ray band.}

\end{deluxetable}

\section{Hardness maps}
\label{sec:app:hardness}

\subsection{Hardness ratios}
\label{sec:app:hardness:ratios}

For counts $H$ and $S$, extracted from two independent energy bands
where $H$ represent the counts in the higher energy (hard) band and $S$
the counts in the lower energy band, the hardness ratio $Q = (H-S)/(H+S)$.
If both $H$ and $S$ are $\ge 0$, then this hardness ratio has
the nice property that $-1 \le Q \le 1$.

The uncertainty in any single measurement of the number of counts
$N$ is $\sigma_{N} \approx 1 + \surd(0.75+N)$, where we have used the
conservative approximation to Poisson errors from \citet{gehrels86}. 
The uncertainty in the hardness ratio $\sigma_{Q}$ is

\begin{equation}
\sigma_{Q} \approx \frac{\sigma_{H+S}}{(H+S)} (1+Q^{2})^{1/2}.
\label{equ:ratio_error}
\end{equation}

In deriving this, we have made use of the fact that for 
normally distributed errors $\sigma_{H-S}^{2} = \sigma^{2}_{H}
+ \sigma_{S}^{2} = \sigma_{H+S}^{2}$.
In practice Poisson statistics should be used when dealing with
X-ray hardness ratios, and this method of combining errors
is not exact. As a robust hardness ratio requires $\ge 20$ counts per band,
the difference between Gaussian and Poisson statistics is
minimal, and we employ the normal method of combination of errors
for convenience.

In practice the counts in both the hard and soft bands must be
modified to subtract the appropriate background counts in each
energy band before making the hardness ratio. The uncertainties
in the hard and soft band counts used in Equ.~\ref{equ:ratio_error}
should take the background subtraction into account.

\subsection{Hardness ratio deviation significance}
\label{sec:app:hardness:deviations}

Hardness maps of diffuse X-ray emission from superwinds are typically
created using smoothed images constructed in the hard and soft energy
bands. Although visually pleasing, it is difficult to assess
the statistical significance of any apparent variation in hardness
seen in such images, in particular at arcminute or sub-arcminute
angular scales.

Fig.~\ref{fig:halo_hardness}b displays a hardness ratio deviation significance
map, which measures
the statistical significance of any local deviation in hardness
ratio from the mean hardness ratio. We define the hardness
ratio deviation significance $\Delta Q/\sigma_{Q}$ at any pixel $i,j$ to be
$(Q_{i,j}-Q_{\rm mean})/\sigma_{Q_{i,j}}$, where $Q_{i,j}$ is the 
hardness ratio at that pixel, $\sigma_{Q_{i,j}}$ is the uncertainty in
the hardness ratio at that point (as given by Equ.~\ref{equ:ratio_error}), 
and $Q_{\rm mean}$ is the mean hardness of the diffuse emission.

High significance deviations ($\Delta Q/\sigma_{Q} \ge 4$) are
likely to indicate real spectral variation. If however,
the hardness deviations are due to noise, then the number of pixels in the 
hardness deviation significance map as a function of 
$\Delta Q/\sigma_{Q}$ should follow a Gaussian distribution
centered on $\Delta Q/\sigma_{Q} = 0$ with a 
FWHM = $2.35 \Delta Q/\sigma_{Q}$. This method provides a convenient method
of quantitatively assessing the significance of spatial variations
in spectral hardness.

\subsection{Diffuse emission hardness ratio deviation significances}

The hardness ratio and hardness ratio deviation significance maps
shown in Figs.~\ref{fig:halo_hardness} \& \ref{fig:disk_hardness} 
apply the method described above to
the diffuse X-ray emission seen by {\it Chandra}.

The two energy bands for the hardness map are chosen based on the
observed diffuse emission spectrum in the region of interest. For
soft thermal spectra observed with the ACIS-S3 chip, the 0.3 -- 0.6 keV
and 0.6 -- 1.0 keV energy bands are perhaps the best choice for
investigating spectral variation.

For each chosen energy band we extract an image of an entire chip,
with all point sources masked out. A background image in the same energy band, 
using the same point source mask, is
also extracted from the background events file.
We do {\em not} interpolate over the holes in the diffuse emission
left by this source subtraction method, but the area affected
is very small given {\it Chandra}'s $\sim$ 1 arcsecond spatial
resolution compared to the 8.3 arcminute square chips.
In order to achieve a meaningful S/N in each pixel after
background subtraction, it is necessary to rebin the raw ACIS
pixels by a factor of 60 or 120 in both RA.~and Dec. The resulting
pixels in the hardness maps are then $29.52$ or $59.04$ arcseconds
on a side.

Each diffuse image is then background subtracted, scaling the background images
by the ratio of the diffuse emission and background total exposure times.
An error image, taking into account both the Poisson noise in the
diffuse image and the background, is also created at this stage.
The background-subtracted hard and soft band images are then combined
to create a hardness map. The error images for the hard and
soft bands are combined with the hardness map to create an error
image for the hardness map.

A mean hardness $Q_{\rm mean}$ is used along with the hardness and hardness
error maps to create the hardness ratio deviation significance map. For
the purposes of investigating whether there is real spatial variation
in the spectral properties of the diffuse emission $Q_{\rm mean}$ should
be the mean hardness ratio over the entire region of interest.



\end{document}